\documentclass[preprint,12pt]{elsarticle}
\usepackage{graphics}
\usepackage{epsfig}
\usepackage{amssymb}

\usepackage{nfontdef}

\newcommand{\ignore}[1]{}

\newcommand{\vek}[1]{\mathchoice{\displaystyle\boldsymbol#1}
{\textstyle\boldsymbol#1}{\scriptstyle\boldsymbol#1}
{\scriptscriptstyle\boldsymbol#1}}
\newcommand{\mat}[1]{\mathchoice{\displaystyle\mathbf#1}
{\textstyle\mathbf#1}{\scriptstyle\mathbf#1}
{\scriptscriptstyle\mathbf#1}}

\newcommand{\di}{\mathrm{d}}

\journal{Elsevier}

\begin{document}

%%%%%%%%%%%%%%%%%%%%%%%%%%%%%%%%%%%%%%%%%%%%%%%%%%%%%%%%%%%%%%%%%
\begin{frontmatter}

\title{Acceleration of Uncertainty Updating in the Description of Transport Processes
in Heterogeneous Materials}

\author[ctu]{Anna Ku\v{c}erov\'a\corref{auth}}
\ead{anicka@cml.fsv.cvut.cz}
\author[ctu]{Jan S\'{y}kora}
\ead{jan.sykora.1@fsv.cvut.cz}
\author[tubs]{Bojana Rosi\'c}
\ead{bojana.rosic@tu-bs.de}
\author[tubs]{Hermann G. Matthies}
\ead{wire@tu-bs.de}
\cortext[auth]{Corresponding author. Tel.:~+420-2-2435-5326;
fax~+420-2-2431-0775}
\address[ctu]{Department of Mechanics, Faculty of Civil Engineering,
  Czech Technical University in Prague, Th\'{a}kurova 7, 166 29 Prague
  6, Czech Republic}
\address[tubs]{Institute of Scientific Computing, Technische Universit\"at
Braunschweig, Hans-Sommer-Str. 65, 38092 Braunschweig, Germany}
%

%%%%%%%%%%%%%%%
\begin{abstract}
  The prediction of thermo-mechanical behaviour of heterogeneous
  materials such as heat and moisture transport is strongly influenced
  by the uncertainty in parameters. Such materials occur e.g.\ in
  historic buildings, and the durability assessment of these therefore
  needs a reliable and probabilistic simulation of transport
  processes, which is related to the suitable identification of
  material parameters.  In order to include expert knowledge as
  well as experimental results, one can employ an updating procedure
  such as Bayesian inference. The classical probabilistic setting of
  the identification process in Bayes's form requires the solution of a
  stochastic forward problem via computationally expensive sampling
  techniques, which makes the method almost impractical.

  In this paper novel stochastic computational techniques such as the
  stochastic Galerkin method are applied in order to accelerate the
  updating procedure. The idea is to replace the computationally
  expensive forward simulation via the conventional finite element
  (FE) method by the evaluation of a polynomial chaos expansion
  (PCE). Such an approximation of the FE model for the forward
  simulation perfectly suits the Bayesian updating.

  The presented uncertainty updating techniques are applied to the
  numerical model of coupled heat and moisture transport in
  heterogeneous materials with spatially varying coefficients defined
  by random fields.
\end{abstract}


\end{frontmatter}
%%%%%%%%%%%%%%%%%%%%%%%%%%%%%%%%%%%%%%%%%%%%%%%%%%%%%%%%%%%%%%%%%
\section{Introduction}
\label{sec:intro}
Durability of structures is influenced by moisture damage processes.
High moisture levels cause metal corrosion, wood decay and other
structural degradation. Thermal expansion and contraction, on the
other hand, can induce large displacements and extensive damage to
structural materials with differing coefficients ,e.g.\ masonry.  The
Charles Bridge in Prague, currently the subject of rehabilitation
works, is a typical example, see \cite{Zeman:ES:2008}.  A study of the
coupled heat and moisture transport behaviour is thus essential in
order to improve the building materials' performance. So far, a vast
number of models have been introduced for the description of transport
phenomena in porous media.  An extensive overview of transport models
can be found in~\cite{Cerny:2002}. In this work we focus on the model
by K\"unzel \cite{Kunzel:1997}, since the predicted results comply
well with the results of experimental measurements
\cite{Sykora:2011:AMC}, once the relevant material parameters are well
estimated.

Material properties are usually determined from experimental
measurements via an identification procedure, see
e.g.~\cite{Kucerova:2007:PHD}.  However, the experimental measurements
as well as the identification methods involve some inevitable
errors. Bayesian updating, employed within this study, provides a
general framework for inference from noisy and limited data.  It
enables mutually involving both expert knowledge of the material, such
as limit values of physical parameters, and information from
experimental observations and measurements.  In other words, it uses
experimental data to update the so-called a priori uncertainty in the
material description and results in a posterior probabilistic
description of material performance \cite{Tarantola:2005}. In
addition, unlike traditional identification techniques that aim to
regularise the ill-posed inverse problem to achieve a point estimate,
the Bayesian identification process leads to a well-posed
problem in an expanded stochastic space.

The main disadvantage of Bayesian updating lies in the significant
computational effort that results from the sampling-based estimation of
posterior densities \cite{Mosegaard:2002:IHEES}.  While
deterministic quadrature or cubature may be attractive alternatives to
Monte Carlo at low to moderate dimensions \cite{Evans:1995:SS},
computationally exhaustive Markov chain Monte Carlo (MCMC) remains the
most general and flexible method for complex and high-dimensional
distributions \cite{Tierney:1994:AS, Gilks:1996}. In a sampling-based
procedure, the posterior distribution must be evaluated for any sample
generated from the prior one in order to decide, whether the sample is
admissible or not. The computation of the posterior involves the
evaluation of the computational model---the FE discretisation of a
non-linear partial differential equation (PDE)---relating model (i.e.\ material)
parameters and observable quantities (i.e.\ model outputs). Hence,
complex and time-consuming models can make the sampling procedure
practically unfeasible.

Bayesian updating of uncertainty in the description of the parameters
of K\"unzel's model is thoroughly described in
\cite{Kucerova:2011:AMC} for the case of heterogeneous material, where
material parameters are described by random fields (RFs). It was shown
that Bayesian updating is applicable even for such a complex and
nonlinear model as K\"unzel's model.  However, the demonstrated
results were performed for a sample with a coarse FE, thereby
rendering the evaluation of the numerical model computationally
relatively cheap. A higher complexity of modelled structure and its
FE-based numerical model lead to time-consuming simulations and are
prohibitive for the sampling procedure. In such a case one may
construct an approximation of the model response and evaluate this
within the sampling procedure in order to render the updating
procedure feasible \cite{Marzouk:2007:JCP, Marzouk:2009:JCP}.

The efficient forward propagation of uncertainty, which may describe
material properties, the geometry of the domain, external loading
etc., from model parameters to model outputs is a main topic of stochastic
mechanics. The recently developed polynomial chaos (PC) variant
of the stochastic finite element method (SFEM)---the spectral SFEM (SSFEM)
\cite{Ghanem:2003, Babuska:2004:SIAM, Matthies:2005:CMAME,
  Matthies:1997:SS, Keese:2004:PHD}---has become one of the promising
techniques in this area. Some of the uncertainties in the model are
represented as random fields/processes.
Here one often employed technique in SFEM computations
is the use of a truncated Karhunen-Lo\`eve
expansion (KLE) to represent the RFs in a computationally
efficient manner by means of a minimal set of random variables (RVs)
\cite{Ghanem:1989, Matthies:2005:CMAME, Matthies:2007:IB,
  Chen:2008:CMAME}, via an eigenvalue decomposition of the covariance.
This approach involves the
introduction of an orthogonal---hence uncorrelated---basis in a space
of RVs.  These are projections of
the RF onto the orthogonal KL eigenfunctions, and in
the case of Gaussian RFs consists of Gaussian RVs.  In that case they
are not only uncorrelated but independent---a computationally very
important property \cite{Keese:2003:TR}.
However, the
material properties very often cannot be modelled as Gaussian due
to crucial constraints such as positive definiteness, boundedness in
some interval, etc. In such a case, one has to adopt non-Gaussian
models and their corresponding approximations, see
\cite{Matthies:1997:SS, Keese:2004:PHD, Colliat:2007:CRM}, often as
a non-linear transformation of a Gaussian RF.  The orthogonal or
uncorrelated RVs alluded to above are not Gaussian in that case, and
hence not independent.  One then may adopt a pure PC representation of
the RF in terms of polynomials of independent Gaussian RVs
\cite{Ghanem:2003}, or---to take advantage of the dimension reduction
inherent in the KLE truncation---one uses the PC representation for
the orthogonal/uncorrelated non-Gaussian RVs from the KLE
\cite{Matthies:2005:CMAME, Matthies:2007:IB}.

In this paper, we focus on K\"{u}nzel's model \cite{Kunzel:1995,
  Kunzel:1997}, defined by uncertain positive-definite material
parameters, modelled as log-normal RFs according to the maximum
entropy principle. Since these RFs are non-Gaussian, their spectral
decomposition (KLE) gives a set of uncorrelated but not necessarily
independent RVs.  To address this problem, we project the RVs onto a
PC basis constructed from Hermite polynomials in independent Gaussian
RVs as alluded to in the previous paragraph.  Such a combined
expansion (KL/PC) is then used to represent the RFs as inputs to the
FE discretisation of the nonlinear K\"{u}nzel model.  The solution
procedure of Galerkin type for this SPDE is chosen in an ``intrusive''
manner based on analytic computations in the PC/Hermite algebra
\cite{Matthies:2007:IB, Rosic:2011:TR, Rosic:2008:JSSCM}. This brings huge
computational savings in case of small and moderate problem
dimensions, but it requires complete knowledge of the model (the FEM
system can not be used in black-box fashion).

Once such a representation is propagated through the physical model,
one obtains a description of all desired output quantities in terms of
simply evaluable functions---in this case polynomials---of known
independent Gaussian RVs.  This is often called a surrogate model or
a response surface.

The paper is organised in the following way. The next
Section~\ref{sec:kunzel} reviews K\"unzel's model.
Section~\ref{sec:het} is focused on the probabilistic description of
heterogeneous material properties where particular material parameters
are not spatially constant. Intrusive stochastic Galerkin method for
computing coefficients of the PC-based surrogate of outputs of
K\"unzel's model is developed in Section~\ref{sec:pce} and the related
outcomes a presented in Section~\ref{sec:res_prop}.  Finally,
Section~\ref{sec:bayes} presents the Bayesian updating procedure on
K\"unzel's model with the results summarized in
Section~\ref{sec:bayes_res}, and Section~\ref{sec:concl} concludes.

\section{Coupled heat and moisture transfer}
\label{sec:kunzel}
K\"{u}nzel \cite{Kunzel:1995, Kunzel:1997} derived balance equations
describing coupled heat and moisture transport through porous media
using the concepts of Krischer and Kiessl. Krischer
\cite{Krischer:1978} identified two transport mechanisms for material
moisture, one being the vapour diffusion and the other being described
as capillary water movement. In other words, he introduced the
gradient of partial pressure in air as a driving force for the water
vapour transport and the gradient of liquid moisture content as the
driving force for the water transport. This model is then extended by
Kiessl \cite{Kiessl:1983} who introduced the so-called moisture
potential $\Phi$ used for unification of the description of moisture
transport in the hygroscopic $\varphi\leq 0.9$ and over-hygroscopic
$\varphi > 0.9$ range (where $\varphi$ is relative humidity). The
introduction of the moisture potential brings several advantages,
especially very simple expressions for the moisture transport across
the interface. On the other hand, the definition of the moisture
potential in the over-hygroscopic range was too artificial, and Kiessl
introduced it without any theoretical background,
see~\cite{Cerny:2002}.

For the description of simultaneous water and water vapour
transport K\"{u}nzel chose the relative humidity $\varphi$ as the
only moisture potential for both the hygroscopic and the over-hygroscopic
range. He also divided the over-hygroscopic region into two
sub-ranges---the capillary water region and supersaturated
region---where different
conditions for water and water vapour transport are considered. In
comparison with Kiessl's or Krischer's model K\"{u}nzel's model
brings certain simplifications. Nevertheless, the proposed model
describes all substantial phenomena and the predicted results
comply well with experimentally obtained data
\cite{Sykora:2011:AMC}. Therefore, it was chosen as a physical basis
for the formulation of the probabilistic framework.

K\"unzel's model is described by the energy balance equation
\begin{equation}
  \frac{\mathrm{d}H}{\mathrm{d}\theta}\frac{\mathrm{d}\theta}{\mathrm{d}t}
  =  \vek{\nabla}^{\mathrm{T}}[\lambda(\varphi)\vek{\nabla} \theta]+h_{\mathrm{v}}(\theta)\vek{\nabla}^{\mathrm{T}}
  [\delta_{\mathrm{p}}(\theta)\vek{\nabla}\{\varphi p_{\mathrm{sat}}(\theta)\}]
\label{eq:TE03}
\end{equation}
and the conservation of mass equation
\begin{equation}
\frac{\mathrm{d}w}{\mathrm{d}\varphi}\frac{\mathrm{d}
\varphi}{\mathrm{d}t}  =  \vek{\nabla}^{\mathrm{T}}
[D_{\varphi}(\varphi)\vek{\nabla}\varphi]+\vek{\nabla}^{\mathrm{T}}
[\delta_{\mathrm{p}}(\theta)\vek{\nabla}\{\varphi p_{\mathrm{sat}}(\theta)\}] \, ,
\label{eq:TE04}
\end{equation}
where the transport coefficients defining the material behaviour are
nonlinear functions of structural responses, i.e.\ the temperature
$\theta [^\circ\mathrm{C}]$ and moisture $\varphi [-]$ fields. We
briefly recall their particular relations \cite{Kunzel:1997}:
\begin{itemize}
\item
Thermal conductivity $[\mathrm{Wm}^{-1}\mathrm{K}^{-1}]$:
\begin{equation}
\lambda = \lambda_0 \left( 1+ \frac{b_{\mathrm{tcs}} w_\mathrm{f} (b-1)
\varphi}{\rho_\mathrm{s} (b - \varphi)} \right) \, . \label{eq_thermcond}
\end{equation}
\item
Evaporation enthalpy of water $[\mathrm{Jkg}^{-1}]$:
\begin{equation}
h_\mathrm{v} = 2.5008 \cdot 10^6 \left( \frac{273.15}{\theta+273.15}
\right)^{(0.267+3.67 \cdot 10^{-4} \theta)} \, .
\end{equation}
\item Water vapour permeability
$[\mathrm{kgm}^{-1}\mathrm{s}^{-1}\mathrm{Pa}^{-1}]$:
\begin{equation}
\delta_\mathrm{p} = \frac{1.9446 \cdot 10^{-12}}{\mu} \cdot \left( \theta +
273.15 \right)^{0.81} \, .
\end{equation}
\item Water vapour saturation pressure $[\mathrm{Pa}]$:
\begin{equation}
p_{\mathrm{sat}} = 611 \exp \left( \frac{17.08 \theta}{234.18 +
\theta} \right) \, .
\end{equation}
\item
Liquid conduction coefficient $[\mathrm{kgm}^{-1}\mathrm{s}^{-1}]$:
\begin{equation}
D_{\varphi} = 3.8 \frac{a^2}{w_\mathrm{f}} \cdot
10^{\frac{3w_\mathrm{f}(b-1)\varphi}{(b-\varphi)(w_\mathrm{f}-1)}} \cdot
\frac{b(b-1)}{(b-\varphi)^2} \, . \label{eq_watvapsat}
\end{equation}
\item
Total enthalpy of building material $[\mathrm{Jm}^{-3}]$:
\begin{equation}
H = \rho_\mathrm{s} c_\mathrm{s} \theta \, . \label{eq_enthalpy}
\end{equation}
\item
Water content $[\mathrm{kgm}^{-3}]$:
\begin{equation}
w = w_\mathrm{f}\frac{(b-1)\varphi}{b-\varphi} \, . \label{eq_watcont}
\end{equation}
\end{itemize}
A more detailed discussion on the transport coefficients can be found in
\cite{Kunzel:1997, Cerny:2009:CMEM}. Some of them
defined by Eqs.~(\ref{eq_thermcond})--(\ref{eq_enthalpy}) depend on
a subset of the material parameters listed in Tab.~\ref{tab_params}.
The approximation factor $b$ appearing in Eqs.~(\ref{eq_thermcond})
and (\ref{eq_watvapsat}) can be determined from the relation:
\begin{equation}
b = \frac{0.8(w_{80}-w_\mathrm{f})}{w_{80}-0.8w_\mathrm{f}} \, ,
\end{equation}
where $w_{80}$ is the equilibrium water content at $0.8\,\mathrm{[-]}$
relative humidity. Moreover, the free water saturation $w_\mathrm{f}$ must
always be greater than $w_{80}$. Therefore we introduce the water content
increment $\mathrm{d}w_\mathrm{f} > 0$ and define the free water saturation
as
\begin{equation}
w_\mathrm{f} = w_{80} + \mathrm{d}w_\mathrm{f} \, .
\end{equation}
Consequently, $w_{80}$ and $\mathrm{d}w_\mathrm{f}$ substitute $b$
and $w_\mathrm{f}$ as material parameters to be identified within
the updating procedure.  Tab.~\ref{tab_params} presents the
resulting list of $W=8$ material parameters to be identified. As
an outcome of such a substitution, all identified parameters
should be positive and thus described by log-normal RFs (a priori
information) with second order statistics (mean values $\mu_q$ and
standard deviations $\sigma_q$) given in Tab.~\ref{tab_params}.
Those particular values are chosen to correspond to materials used
in masonry \cite{Pavlik:SFR:2010}.
\begin{table}[h!]
\centering
\begin{tabular}{lllcc}
\multicolumn{3}{l}{Parameter} & $\mu_q$ & $\sigma_q$ \\
\hline
$\mathrm{d}w_\mathrm{f}$ & $\mathrm{[kgm^{-3}]}$ &  water content
increment & 100 & 20 \\
$w_{\mathrm{80}}$ & $\mathrm{[kgm^{-3}]}$ &  water content at $0.8\,\mathrm{[-]}$ relative humidity & 50 & 10 \\
$\lambda_{\mathrm{0}}$ & $\mathrm{[Wm^{-1}K^{-1}]}$ &  thermal conductivity of dry material & 0.3 & 0.1 \\
$b_{\mathrm{tcs}}$ & $\mathrm{[-]}$ &  thermal conductivity supplement & 10 & 2  \\
$\mu$ & $\mathrm{[-]}$ &  water vapour diffusion resistance factor & 12 & 5 \\
$a$ & $\mathrm{[kgm^{-2}s^{-0.5}]}$ &  water absorption
coefficient & 0.6 & 0.2 \\
$c_\mathrm{s}$ & $\mathrm{[Jkg^{-1}K^{-1}]}$ &  specific heat capacity & 900 & 100 \\
$\rho_\mathrm{s}$ & $\mathrm{[kgm^{-3}]}$ &  bulk density of building material & 1650 & 50 \\
\hline
\end{tabular}
\caption{Mean values and standard deviations of
  material parameters} \label{tab_params}
\end{table}

The partial differential equations
(\ref{eq:TE03}) and (\ref{eq:TE04}) are discretised in space by
standard finite elements.  This also goes well with the
use of the stochastic Galerkin method for the discretisation
in the stochastic space. Performing first only the spatial
discretisation, the
temperature and moisture fields are spatially approximated as
\begin{equation}
\theta(\vek{x})= \sum_{n=1}^N \phi_n(\vek{x})u_{\theta,n}, \qquad
\varphi(\vek{x})= \sum_{n=1}^N \phi_n(\vek{x})u_{\varphi,n}
\label{eq:DDE3}
\end{equation}
where $N$ is the number of nodes in FE discretisation,
$\phi_n(\vek{x})$ are the shape functions (according to the type of
used elements) and $u_{\theta,n}$ and $u_{\varphi,n}$ are the nodal
values of temperature field $\theta$ and moisture field $\varphi$,
respectively.

Using the approximations Eq.~\eqref{eq:DDE3} and Eqs.~\eqref{eq:TE03},
\eqref{eq:TE04}, we obtain a set of first order differential equations
\begin{equation}
  \vek{K}(\vek{u})\vek{u}+\vek{C}(\vek{u})\frac{\mathrm{d}\vek{u}}{\mathrm{d}t} =  \vek{F},\label{eq:MS1}
\end{equation}
where $\vek{K}(\vek{u})$ is the conductivity matrix,
$\vek{C}(\vek{u})$ is the capacity matrix,
$\vek{u}^\mathrm{T}=(u_{\theta,1}, \dots,
u_{\theta,N},u_{\varphi,1},\dots,u_{\varphi,N})$ is the vector of
nodal values, and $\vek{F}$ is the vector of prescribed fluxes
transformed into nodes. For a detailed formulation of the matrices
$\vek{K}(\vek{u})$ and $\vek{C}(\vek{u})$ and the vector
$\vek{F}$, we refer the interested reader to the doctoral thesis
\cite[Chapter 3.1]{Sykora:2010:phd}.

The numerical solution of the system
Eq.~\eqref{eq:MS1} is based on a simple temporal finite difference
discretisation.  If we use time steps $\Delta\tau$ and denote the
quantities at time step $i$ with a corresponding superscript,
the time-stepping equation is
\begin{equation}
\vek{u}^{i+1} = \vek{u}^{i} + \Delta\tau [(1-\gamma)\dot{\vek{u}}^{i}
+ \gamma\dot{\vek{u}}^{i+1}], \label{eq:MS4}
\end{equation}
where $\gamma$ is a generalised midpoint integration rule
parameter. In the results presented in this paper the
Crank-Nicolson (trapezoidal rule) integration scheme with $\gamma
= 0.5$ was used.  Expressing $\dot{\vek{u}}^{i+1}$ from
Eq.~\eqref{eq:MS4} and substituting into the Eq.~\eqref{eq:MS1},
one obtains a system of non-linear equations:
\begin{equation}
  (\gamma \Delta \tau \vek{K}^{i+1}  + \vek{C}^{i+1}) \vek{u}^{i+1}
  = \gamma \Delta \tau \vek{F}^{i+1}  + \vek{C}^{i+1} [\vek{u}^{i}
  + \Delta \tau (1-\gamma) \dot{\vek{u}}^{i}],\label{eq:MS5}
\end{equation}
which can be solved by some iterative method such as Newton-Raphson.
For clarification and easier reading, we rewrite Eq.~\eqref{eq:MS5}
using the symbols
$\vek{A}^{i+1}(\vek{u}^{i+1}) := \gamma\Delta\tau\vek{K}^{i+1}(\vek{u}^{i+1})
 + \vek{C}^{i+1}(\vek{u}^{i+1})$ and $\vek{f}^{i+1}(\vek{u}^{i+1}) :=
 \gamma\Delta\tau\vek{F}^{i+1} + \vek{C}^{i+1}(\vek{u}^{i+1})[\vek{u}^i +
\Delta\tau(1-\gamma)\dot{\vek{u}}^i]$ in the following form
\begin{equation}
\vek{A}^{i+1}(\vek{u}^{i+1})\vek{u}^{i+1}
 = \vek{f}^{i+1}(\vek{u}^{i+1}).\label{eq:MS6}
\end{equation}

\section{Uncertain properties of heterogeneous materials}
\label{sec:het}
When dealing with heterogeneous material, some material parameters can
vary spatially in an uncertain fashion and therefore RFs are
suitable for their description. This means that the uncertainty in
a particular material parameter $q$ is modelled by defining
$q(\vek{x})$ for each $\vek{x} \in \C{G}$ as a RV $q(\vek{x}):
\mit{\Omega} \rightarrow \D{R}$ on a suitable probability space
$(\mit{\Omega} ,\mathscr{S},\D{P})$ in some bounded admissible region
$\C{G} \subset \D{R}^d$. As a consequence, $q : \C{G} \times
\mit{\Omega} \rightarrow \D{R}$ is a RF and one may identify
$\mit{\Omega}$ with the set of all possible realisations of $q$.
Alternatively,
$q(\vek{x},\omega)$ can be seen as a collection of real-valued RVs
indexed by $\vek{x} \in \C{G}$.

The description of log-normal RFs given in Tab.~\ref{tab_params} can
be derived from a Gaussian RF $g(\vek{x}, \omega)$, which is defined
by its mean
\begin{equation}
\mu_g(\vek{x}) = \D{E}[g(\vek{x},\omega)] = \int_{\mit{\Omega}}
g(\vek{x},\omega) \, \D{P}(\di \omega)\, ,
\end{equation}
and its covariance
\begin{eqnarray}
C_g(\vek{x},\vek{x}^{\prime}) & = & \D{E}[(g(\vek{x},\omega) -
\mu_g(\vek{x}))
(g(\vek{x}^{\prime},\omega) - \mu_g(\vek{x}^{\prime}))] \nonumber \\
& = & \int_{\mit{\Omega}} (g(\vek{x},\omega) - \mu_g(\vek{x}))
(g(\vek{x}^{\prime},\omega) - \mu_g(\vek{x}^{\prime})) \,
\D{P}(\di \omega) \, .
\end{eqnarray}
The log-normal RF $q(\vek{x},\omega)$ can be then obtained by a
nonlinear transformation of a zero-mean unit-variance Gaussian RF
$g(\vek{x},\omega)$ \cite{Matthies:2007:IB, Rosic:2008:JSSCM} as
\begin{equation}
  q(\vek{x},\omega) = \exp ( \mu_g(\vek{x}) + \sigma_g g(\vek{x},\omega)
  ) \, .
\label{eq_logfield}
\end{equation}
The statistical moments $\mu_g$ and $\sigma_g$ of the Gaussian field
can be obtained from the statistical moments $\mu_q$ and $\sigma_q$
given for the log-normally distributed material property according to
the following relations \cite{Rosic:2008:JSSCM}:
\begin{equation}
\sigma_g^2  =  \ln \left( 1 + \left( \frac{\sigma_q}{\mu_q}
\right)^2 \right) \, , \quad \mu_g  =  \ln \mu_q - \frac{1}{2}
\sigma_g^2 \, \label{eq_sigmag}.
\end{equation}

In numerical computation random fields are first spatially discretised
by finite element method (see Eq.~\eqref{eq:DDE3}) into a finite
collection of points $\{{\vek{x}}^n_{i=1}\} \in \C{G}$. Further, the
semi-discretised RF are described by a finite---but probably very
large---number of RVs $\vek{q}(\omega) = (q(\vek{x}_1,\omega), \dots,
q(\vek{x}_n,\omega))$, which are usually highly correlated. Large
number of RVs is, however, very challenging for the efficient
numerical implementation of forward problem, as well as for MCMC
identification. As already alluded to previously, the number of RVs
can be reduced by the approximation $\hat{\vek{g}}(\omega)$ of a RF
$\vek{g}(\omega)$ based on a truncated KLE including much smaller
number of RVs \cite{Matthies:2007:IB, Marzouk:2009:JCP}.  Here we use
the KLE on the underlying Gaussian field $\vek{g}(\omega)$, and hence
the RVs in the KLE are independent Gaussian RVs, as already indicated
above.

The spatial discretisation of a given RF concerns also the discretisation
of corresponding covariance function
$C_g(\vek{x},\vek{x}^{\prime})$ into the covariance matrix
$\vek{C}_g$ which is symmetric and positive definite
\cite{Matthies:2005:CMAME, Matthies:2007:IB}. The KLE is based on the spectral
decomposition of the covariance matrix $\vek{C}_g$ leading to the
solution of a symmetric matrix eigenvalue problem
\begin{equation}
  \vek{C}_g \vek{\psi}_i = \varsigma_i^2 \vek{\psi}_i \, ,
\label{eq_fredholm}
\end{equation}
where $\vek{\psi}_i$ are orthogonal eigenvectors and $\varsigma_i^2$ are
positive eigenvalues ordered in a descending order. The KLE approximation
$\hat{\vek{g}}(\omega)$ of a RF $\vek{g}(\omega)$
can then be written as
\begin{equation}
\hat{\vek{g}}(\omega) =
\vek{\mu}_g+\sum_{i=0}^{M} \varsigma_{i} \, \xi_{i}(\omega) \vek{\psi}_{i},
\label{eq_kleserie}
\end{equation}
where $\xi_{i}(\omega) = \vek{\psi_i}^T (\vek{g}(\omega) -
\vek{\mu}_g)/\varsigma_i$ are uncorrelated RVs of zero mean and unit
variance, and in case that $g(\vek{x}, \omega)$ and hence
$\vek{g}(\omega)$ are Gaussian, then $x_i(\omega)$ are Gaussian and
independent. The number $M \leq n$---the number of points used for the
discretisation of the spatial domain---is chosen such that
Eq.~\eqref{eq_kleserie} gives a good approximation, i.e.\ captures a
high proportion of the total variance.
Higher values of $M$ lead to better description of a RF, smaller
values imply faster exploration by MCMC. The eigenvalue problem
Eq.~\eqref{eq_fredholm} is usually solved by a Krylov subspace method with
a sparse matrix approximation. For large eigenvalue problems, the
authors in \cite{Khoromskij:2008} propose  efficient low-rank and
data sparse hierarchical matrix techniques. The approximation of a
non-Gaussian RF can be then obtained by a nonlinear transformation of
the KLE obtained for a Gaussian RF such as in our particular case,
where the approximation of a given RF $\hat{\vek{q}}(\omega)$ is
obtained from the Eq.~\eqref{eq_logfield} by the substitution of the
Gaussian RF $\vek{g}(\omega)$ by its KLE $\hat{\vek{g}}(\omega)$.

We assume full spatial correlation among material properties,
i.e.\ spatial fluctuations for all parameters differ only in
magnitude. Taking into account a log-normal distribution of the
parameters, the final formulation of the RF describing the
parameter $q$ then becomes
\begin{equation}
  \hat{\vek{q}}(\omega) = \exp \left( \vek{\mu}_g +
\sigma_g \sum_{i=1}^{M} \sqrt{\varsigma_i}\xi_i(\omega) \vek{\psi}_i
  \right) \, ,
\label{eq_logfield_ext}
\end{equation}
where the exponential is to be used at each spatial point, i.e.\
for each component of the vector inside the parentheses. The
statistical moments $\vek{\mu}_g$ and $\sigma_g$ are derived
  from the prior mean $\vek{\mu}_q$ and standard deviation $\sigma_q$
  for each material parameter according to Eq.~(\ref{eq_sigmag}). The
  eigenvectors $\vek{\psi}_i$ are obtained for the a priori
  exponential covariance function
\begin{equation}
C(\vek{x}, \vek{x}^{\prime}) = \exp \left( -\frac{|r_1|}{l_{x_{1}}} -
\frac{|r_2|}{l_{x_{2}}} \right)\, ,
\label{eq_normkern}
\end{equation}
where $\vek{r} = (r_1, r_2) = \vek{x} - \vek{x}^{\prime}$, and
$l_{x_{1}} = 0.1$~$\mathrm{[m]}$ and $l_{x_{2}} =
0.04$~$\mathrm{[m]}$ are a priori covariance lengths.
Determination of correlation lengths is generally not obvious. In
material modelling, one possible way is based on image analysis as
described in \cite{Lombardo:2009:IJMCE}. A numerical study for a
differing number of modes $M$ included in the KLE is presented in
\cite{Kucerova:2011:AMC}.

\section{Surrogate of K\"unzel's model}
\label{sec:pce}
While the KLE can be efficiently applied to reduce the number of RVs and
thus to accelerate the exploration of the MCMC method in terms of the
number of samples, construction of a surrogate of the computational
model can be used for a significant acceleration of each sample
evaluation. In \cite{Marzouk:2007:JCP, Marzouk:2009:JCP}
methods  were introduced for accelerating Bayesian inference in this
context through the use of stochastic spectral methods to propagate
the prior
uncertainty through the forward problem. Here we employ the
stochastic Galerkin method
\cite{Babuska:2004:SIAM, Matthies:2005:CMAME} to construct the
surrogate of K\"unzel's model based on polynomial chaos expansion
(PCE).

According to Eq.~(\ref{eq_logfield_ext}), all model parameters are
characterised by $M$ independent standard Gaussian RVs
$\vek{\xi}(\omega) = [\xi_1(\omega), \dots, \xi_{M}(\omega)]$. Hence,
the discretised model response $\vek{u}(\vek{\xi}(\omega))= \left(
  \dots, u_i(\vek{\xi}(\omega)) \dots \right)^{\mrm{T}}$ is a random
vector which can be expressed in terms of the same RVs
$\vek{\xi}(\omega)$.  Since $\vek{\xi}(\omega)$ are independent
standard Gaussian RVs, Wiener's PCE based on multivariate Hermite
polynomials---orthogonal in the Gaussian
measure---$\{H_\alpha(\vek{\xi}(\omega))\}_{\alpha \in \C{J}}$ (see
\cite{Matthies:2005:CMAME, Matthies:2007:IB} for the notation) is the
most suitable choice for the approximation
$\tilde{\vek{u}}(\vek{\xi}(\omega))$ of the model response
$\vek{u}(\vek{\xi}(\omega))$ \cite{Xiu:2002:SIAM}, and it can be
written as
\begin{equation}
\tilde{\vek{u}}(\vek{\xi}(\omega)) = \sum_{\alpha \in \C{J}} \vek{u}_\alpha
 H_{\alpha}(\vek{\xi}(\omega)) \,
\label{eq_Tapprox}
\end{equation}
where $\vek{u}_\alpha$ is a vector of PC coefficients and the index set $\C{J}
\subset \D{N}_0^{(\D{N})}$ is a finite set of non-negative integer
sequences with only finitely many non-zero terms, i.e.\ multi-indices,
with cardinality $|\C{J}| = R$.
We collect all the PC coefficients in $\mat{u} := [\dots,
\vek{u}_\alpha, \dots]_{\alpha \in \C{J}}$.
Assuming the uncertainty in all material parameters listed in
Tab.~\ref{tab_params} and consequently in the model response,
Eq.~\eqref{eq:MS6} can be rewritten as
\begin{equation}
  \vek{A}^{i+1}\left(\vek{\xi};\vek{u}^{i+1}(\vek{\xi})\right)
  \vek{u}^{i+1}(\vek{\xi})
  = \vek{f}^{i+1}\left(\vek{\xi};\vek{u}^{i+1}(\vek{\xi})\right).
  \label{eq:MS6omega}
\end{equation}
Substituting the model response $\vek{u}^{i+1}(\vek{\xi})$ by its PC
approximation $\tilde{\vek{u}}^{i+1}(\vek{\xi})$ given in
Eq.~\eqref{eq_Tapprox} and applying a Bubnov-Galerkin projection, one
requires that the weighted residuals vanish:
\begin{equation}
  \forall \beta \in \C{J} : \quad \D{E}([
  \vek{f}^{i+1}\left(\vek{\xi};\tilde{\vek{u}}^{i+1}(\vek{\xi})\right)
  - \tilde{\vek{A}}^{i+1}(\vek{\xi})
  \tilde{\vek{u}}^{i+1}(\vek{\xi})]H_\beta(\vek{\xi}))=0 \, ,
\label{eq:bubgal}
\end{equation}
where $\tilde{\vek{A}}^{i+1}(\vek{\xi}) :=
\vek{A}^{i+1}\left(\vek{\xi};\tilde{\vek{u}}^{i+1}(\vek{\xi})\right)$.
Eq.~\eqref{eq:bubgal} together with Eq.~\eqref{eq_Tapprox} leads to
\begin{equation}
  \forall \beta \in \C{J} : \quad \sum_{\alpha \in \C{J}}
  \D{E}\left(H_\beta(\vek{\xi}) \tilde{\vek{A}}^{i+1}(\vek{\xi})
    H_\alpha(\vek{\xi})\right) \vek{u}_\alpha^{i+1} = \D{E}(
  \vek{f}^{i+1}(\vek{\xi}) H_\beta(\vek{\xi})) \, ,
\label{eq:PCsystem}
\end{equation}
which is a non-linear system of equations of size $N \times R$.

The approximation $\tilde{\vek{u}}^{i+1}(\vek{\xi})$ can be
represented through its PC coefficients $\mat{u}^{i+1}$, and similarly
for all other quantities. Denoting the block-matrix
$\mat{A}^{i+1}(\mat{u}^{i+1}) := \left(\D{E}\left(H_\beta(\vek{\xi})
\vek{A}^{i+1}(\vek{\xi}) H_\alpha(\vek{\xi})\right)\right)_{\beta, \alpha \in \C{J}}$,
and the right hand side $\mat{f}^{i+1} :=
(\D{E}(\vek{f}^{i+1}(\vek{\xi}) H_\beta(\vek{\xi})))_{\beta\in\C{J}}$, the system
Eq.~\eqref{eq:PCsystem} may succinctly be written as
\begin{equation}
  \label{eq:PCsyst_mat}
  \mat{A}^{i+1}(\mat{u}^{i+1}) \mat{u}^{i+1} = \mat{f}^{i+1}.
\end{equation}
The matrix $\mat{A}^{i+1}$ has more structure than is displayed here,
but this is outside the scope of this paper;
see \cite{Matthies:2005:CMAME, Matthies:2007:IB} for details and
possible computational procedures.

The evaluation of expected values in Eq.~\eqref{eq:PCsystem} can often
be performed analytically in intrusive Galerkin procedures---that is
their advantage---using the Hermite algebra \cite{Matthies:2007:IB}.
In case they are to be computed numerically, they may be
approximated by a weighted sum of samples drawn from the
prior distributions. To that purpose, one can apply some
integration technique: the Monte Carlo (MC) method, the quasi-Monte Carlo
(QMC) method, or some quadrature rule, see \cite{Matthies:2007:IB} for
a recent review. The latter ones allow to take advantage of a possibly
regular behaviour
in the stochastic variables and consequently reduce the number of
samples. Since the system of equations Eq.~\eqref{eq:PCsystem} can be
quite large, the evaluation of the left hand side for each sample of
$\vek{\xi}$ becomes costly. Here we apply a sparse-grid Smolyak quadrature
rule \cite{Smolyak:1963, Keese:2003:TR, Matthies:2005:CMAME,
  Matthies:2007:IB}, sometimes also named hyperbolic cross integration
method, which is an efficient alternative for integration over Gaussian RVs.

After solving the system Eq.~\eqref{eq:PCsyst_mat}, one has via
Eq.~\eqref{eq_Tapprox} a surrogate representation of the model
outputs. This model approximation may be evaluated orders of magnitude
more quickly than the evaluation containing the full FE simulation.

\section{Numerical results for the uncertainty propagation}
\label{sec:res_prop}
\begin{figure}[h!]
\centering
\includegraphics[keepaspectratio,width=10cm]{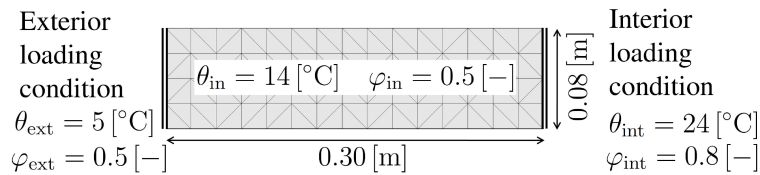}
\caption{Experimental setup} \label{fig_scheme}
\end{figure}

For an illustration of the described method, we employ the same simple
example as in \cite{Kucerova:2011:AMC} with the two-dimensional
rectangular domain discretised by an FE mesh into $N = 80$ nodes and
$120$ triangular elements. Its geometry together with the specific
loading conditions are shown in Fig.~\ref{fig_scheme}.
The initial temperature is $\theta_{\mathrm{in}} = 14$
[$^{\circ}$C] and the moisture $\varphi_{\mathrm{in}} = 0.5$ [-]
in the whole domain. One side of the domain is submitted to
exterior loading conditions $\theta_{\mathrm{ext}} = 5$
[$^{\circ}$C] and $\varphi_{\mathrm{ext}} = 0.5$ [-], while the
opposite side is submitted to interior loading conditions
$\theta_{\mathrm{int}} = 24$ [$^{\circ}$C] and
$\varphi_{\mathrm{int}} = 0.8$ [-]. The solution of the
time-dependent problem in Eq.~\eqref{eq:PCsyst_mat} also involves
a discretisation of the time domain $\C{T}$ into $T = 151$ time
steps and hence the PCE-based surrogate model consist of $N \times
T = 12,080$ PCEs for the temperature, and the same for the
moisture.

In order to describe the accuracy of such a surrogate model, let
us define the MC estimate of the error expectation
$\varepsilon(\vek{u})$ as a relative difference between two
response fields $\vek{u}^a$ and $\vek{u}^b$ over the discretised
spatial and time domain as
\begin{equation}
  \varepsilon%_{\C{G} \otimes \C{T}}
  (\vek{u}) := \D{E}_{\mit{\Omega}}
  \left( \sum_{n=1}^N \sum_{t=1}^T
  \frac{\left|u^a_{n,t} - u^b_{n,t}\right|}{u^a_{n,t}} \right) \, .
\end{equation}

The quality of a PC-based surrogate model depends on the number
$M$ of eigenmodes involved in KLE describing the fields of
material properties as well as on the degree of polynomials $P$
used in the expansion Eq.~\eqref{eq_Tapprox}\footnote{We assume
the full PC expansion,
  where number of polynomials $R$ is fully determined by the degree of
  polynomials $P$ and number of eigenmodes $M$ according to the
  well-known relation $R = (M+P)!/(M!P!)$.}.
\begin{figure} [ht!]
\centering
\begin{tabular}{cc}
\includegraphics*[width=65mm,keepaspectratio]{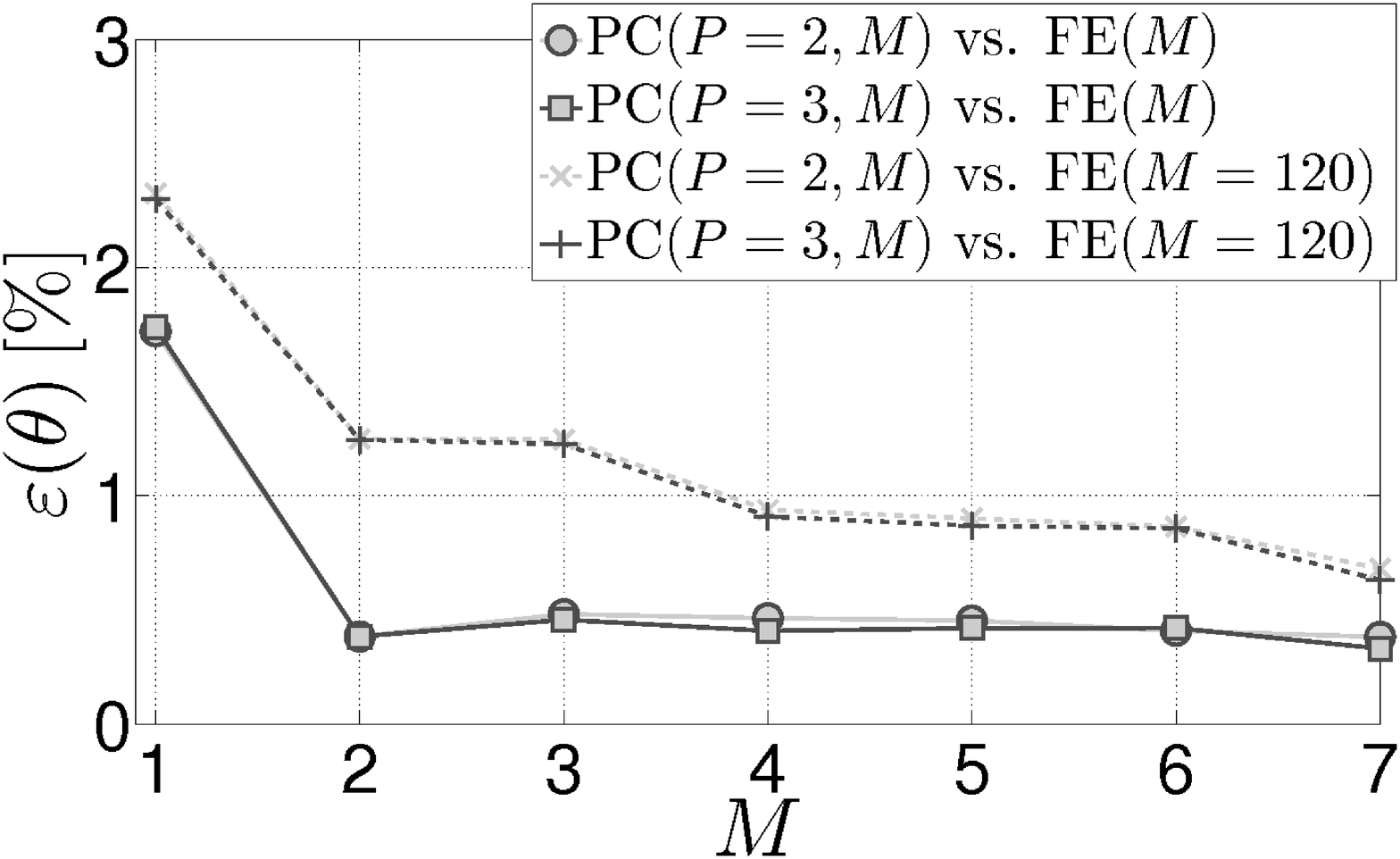}&
\includegraphics*[width=65mm,keepaspectratio]{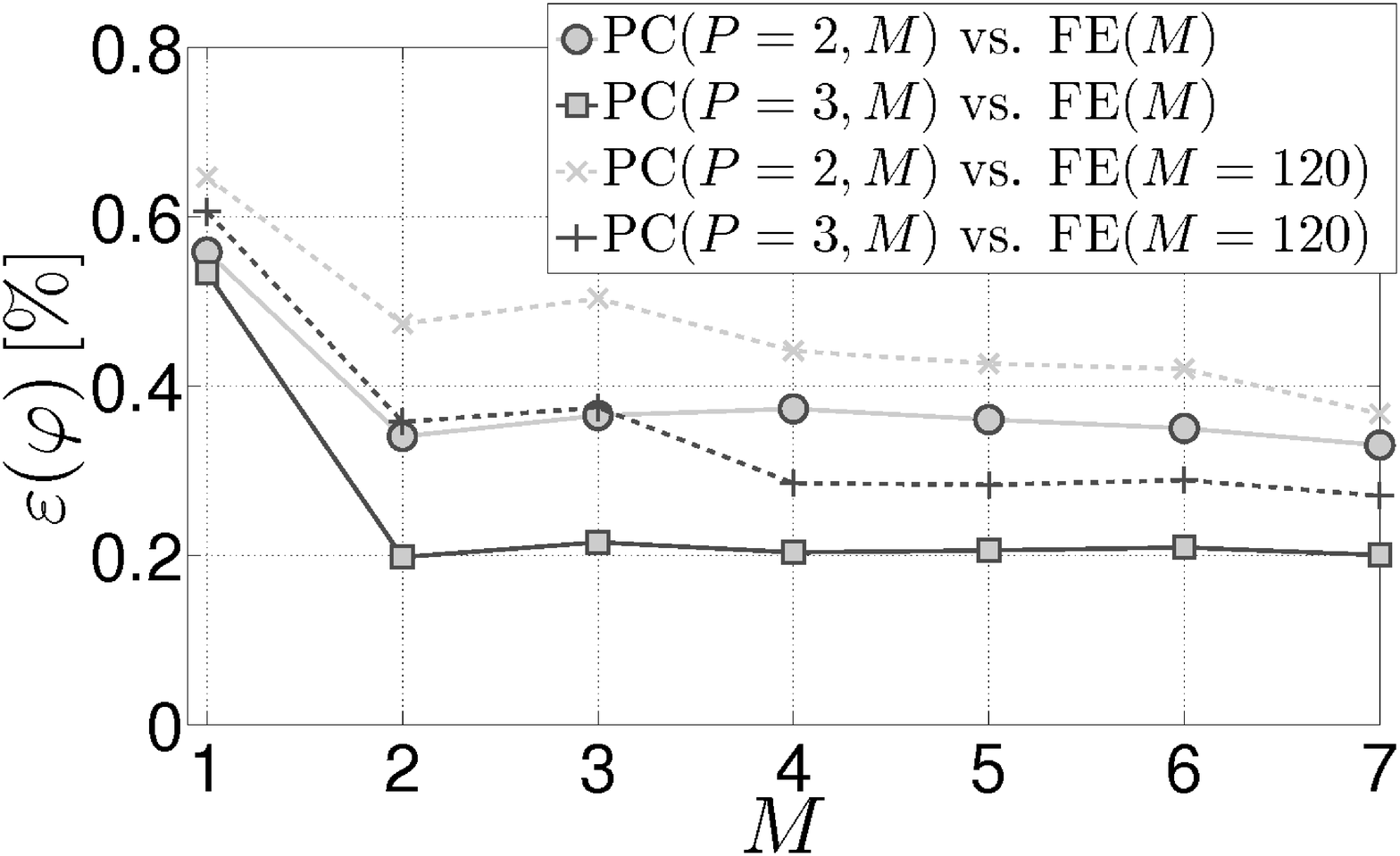}\\
(a)&(b)
\end{tabular}
\caption{ Errors in approximation of the temperature (a) and the
  moisture (b) field induced by PCE and KLE as a function of number of
  eigenmodes.}
\label{fig_wholeerr}
\end{figure}
Figure~\ref{fig_wholeerr} shows the error estimate
$\varepsilon(\vek{u})$ computed for different numbers of
eigenmodes $M$ and for the polynomial order $P=2$ and $P=3$. Here,
the response fields $\vek{u}^a$ are
  computed by the FEM based on one realization of the KLE of the
  parameter fields (further shortly called FE simulations) and the
  response fields $\vek{u}^b$ are obtained by evaluation of the
  constructed PCE in the same sample point. In order to distinguish
  the portion of error induced by the KL approximation of the parameter
  fields, the estimate $\varepsilon(\vek{u})$ is computed once for the FE
  simulations using all $M=120$ (dashed lines), and once for the FE
  simulation using the same number of eigenmodes as in the constructed
  PCE (solid lines). In other words, the solid lines represent the
error induced by PC approximation and the difference between the
solid and corresponding dashed line quantifies the error induced
by the KL approximation of the parameter fields.

\begin{figure} [ht!]
\centering
\begin{tabular}{cc}
\includegraphics*[width=65mm,keepaspectratio]{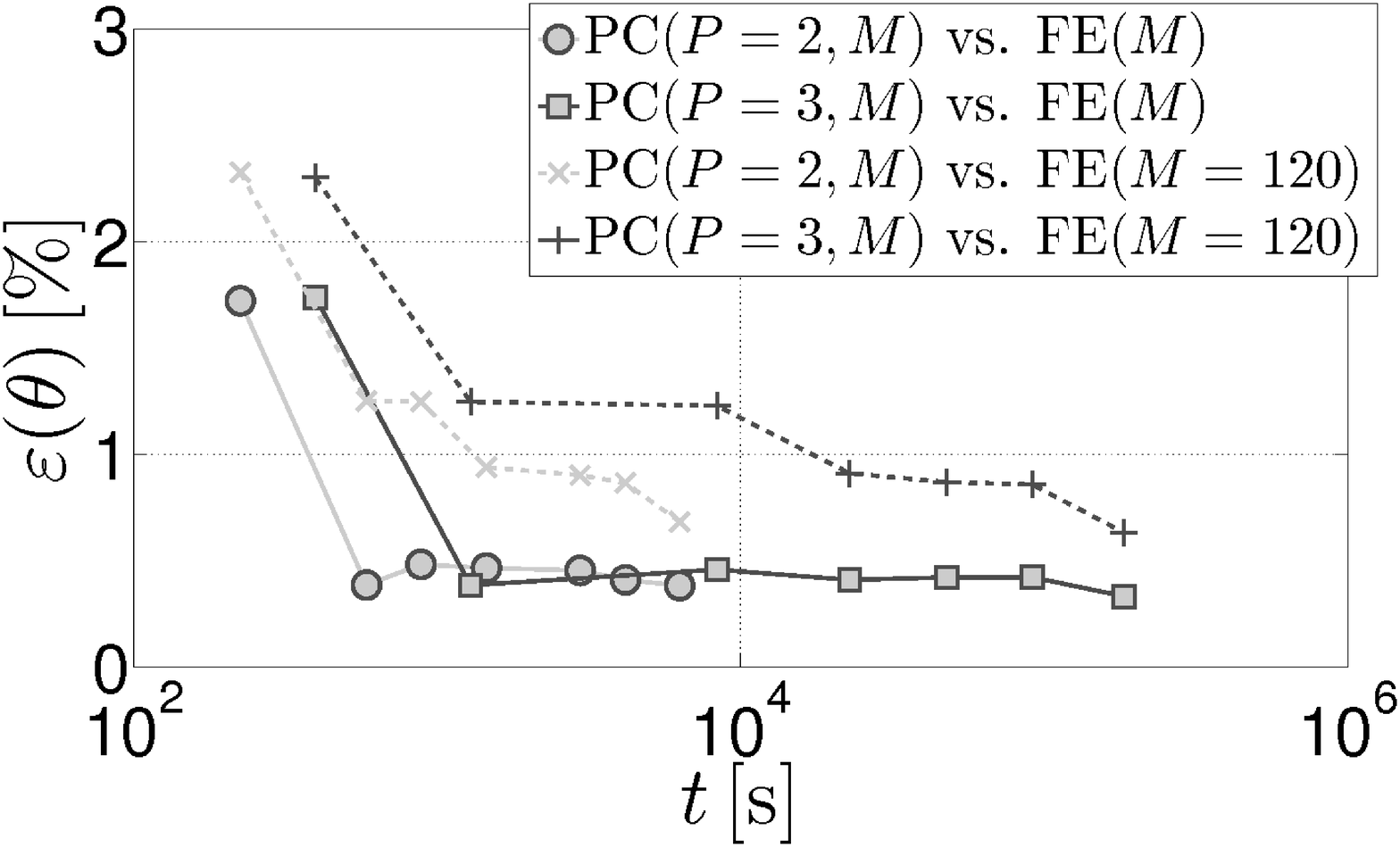}&
\includegraphics*[width=65mm,keepaspectratio]{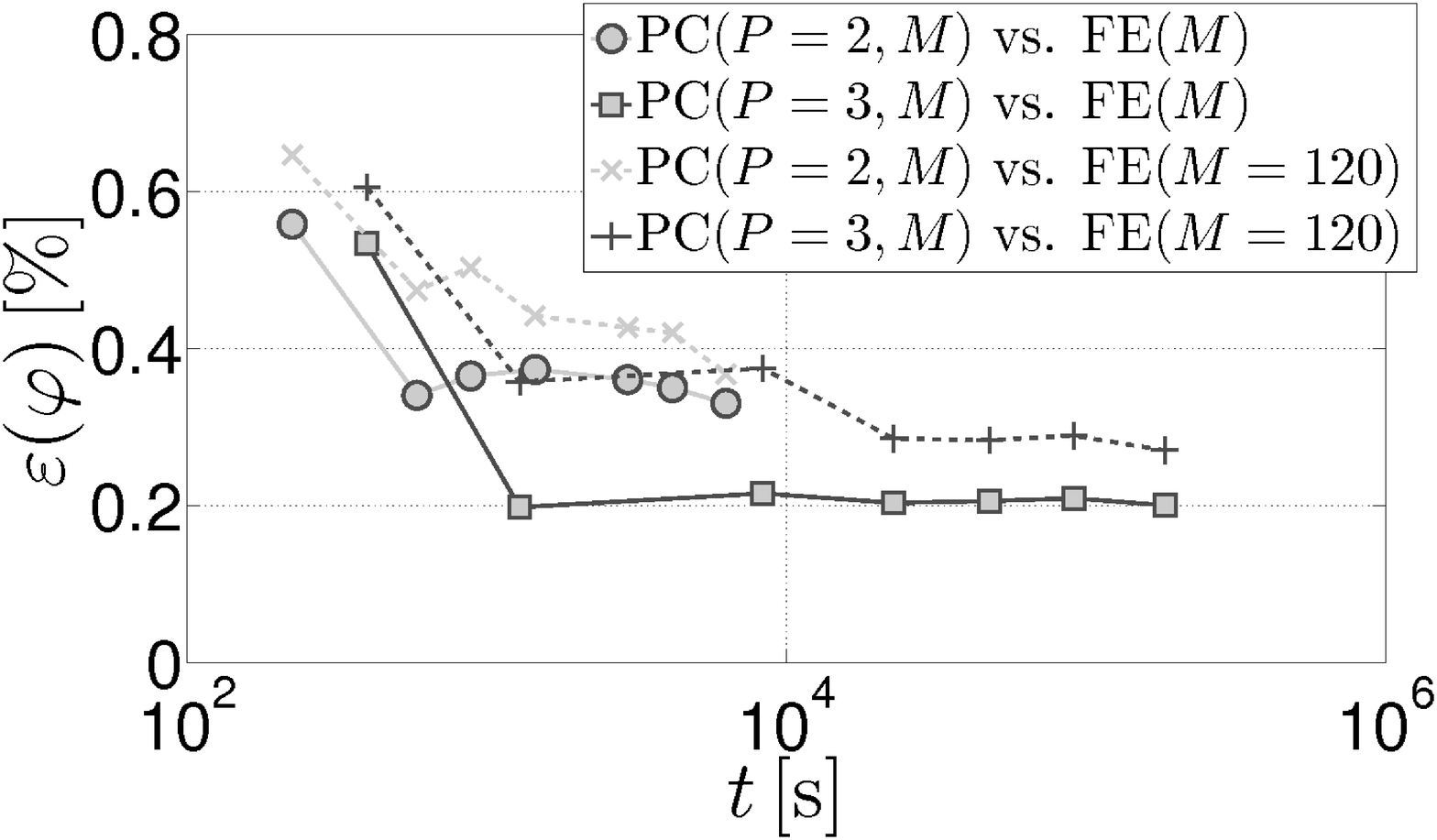}\\
(a)&(b)
\end{tabular}
\caption{ Errors in approximation of the temperature (a) and the
  moisture (b) field induced by PCE and KLE as a function of
  computational time needed for a PCE construction.}
\label{fig_wholeerr_time}
\end{figure}
Figure~\ref{fig_wholeerr_time} represents the same errors
$\varepsilon(\vek{u})$ as Fig.~\ref{fig_wholeerr}, but this time
with respect to the computational effort needed for the
computation of PC coefficients. Regarding the obtained results, we
focus our following computations on the KL approximation of the
material parameters including $M=7$ eigenmodes and a PCE of order
$P=2$ providing, at reasonable time, sufficiently good
approximation of the model response, namely of the temperature
field where the errors are more significant.

\begin{figure} [ht!]
\centering
\begin{tabular}{cc}
\includegraphics*[width=65mm,keepaspectratio]{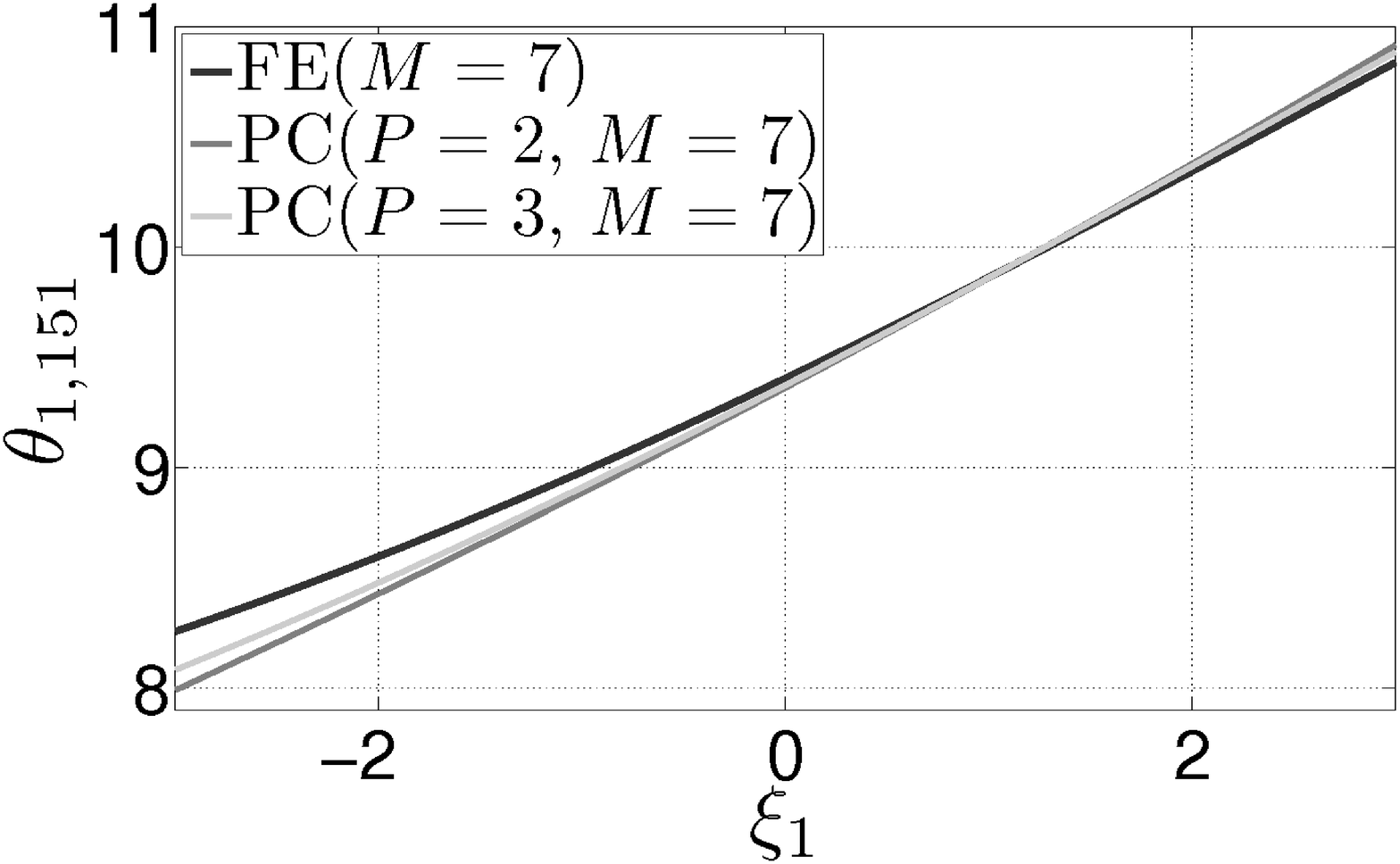}&
\includegraphics*[width=65mm,keepaspectratio]{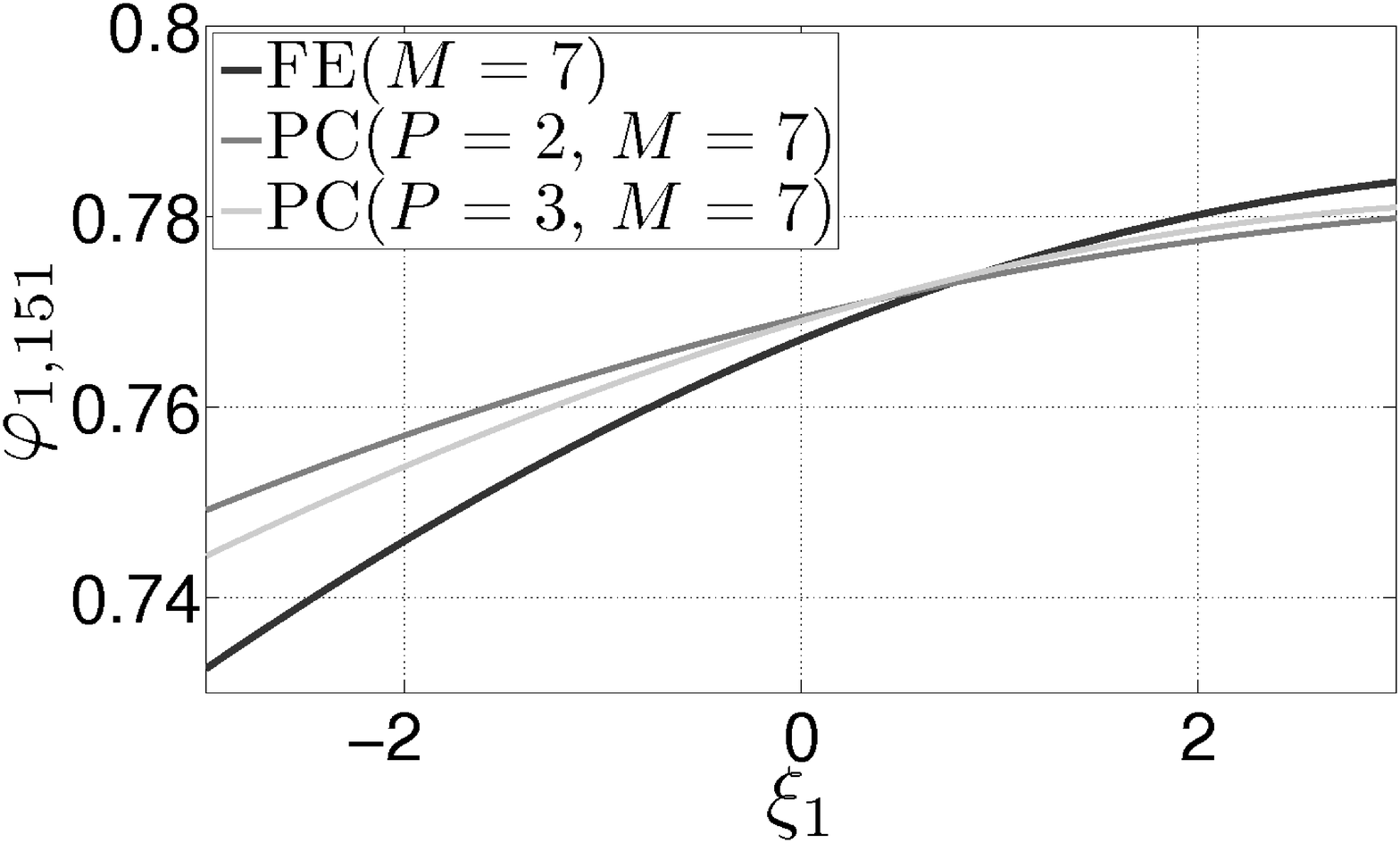}\\
(a)&(b)
\end{tabular}
\caption{Detailed comparison of the temperature (a) and moisture (b)
  with their PC approximation as functions of the first stochastic variable
  $\xi_1$.}
\label{fig_pceerr_detail}
\end{figure}

For a more detailed presentation of the PCE accuracy,
Fig.~\ref{fig_pceerr_detail} compares the model response in one node
of FE mesh (the node No. $1$ at Fig.~\ref{fig_observations}) at the
time $t = 400 [\mathrm{h}]$ obtained by the FE simulation and by the
PCE as a function of the first stochastic variable $\xi_1$.

\section{Bayesian updating procedure}
\label{sec:bayes}

In the Bayesian approach to parameter identification, we assume
three sources of information and uncertainties which should be
taken into account. The first one is the prior knowledge about the
model/material parameters $q_i(\omega)$ defining the prior density
functions. In our particular case, we know that all the identified
parameters are positive-definite and the log-normal random fields
with the statistical moments given in Tab.~\ref{tab_params} are
suitable for a description of the prior information. In fact, they
are maximum entropy distributions for this case. We describe the
material parameters using the KLE which is fully defined by a
finite set of standard Gaussian variables $\vek{\xi} =
[\xi_1(\omega), \dots, \xi_M(\omega)]$ with the probability
density function (pdf) $p_{\vek{\xi}}(\vek{\xi})$ and thus, the
updating procedure can be performed in terms of $\vek{\xi}$
turning them into non-Gaussian variables.

Other source of information comes from measurements, which are
violated by uncertain experimental errors
$\epsilon(\bar{\omega})$. Last uncertainty $\bar{\bar{\omega}}$
arises from imperfection of the numerical model, when for example
the description of a real system does not include all important
phenomena.  However, it is a common situation that the
imperfection of the system description cannot be distinguished
from measurement error $\epsilon$ and the modelling uncertainties
$\bar{\bar{\omega}}$ can be hidden inside the measuring error
$\epsilon(\bar{\omega})$. Then we can define the pdf
$p_{\vek{z}}(\vek{z})$ for noisy measurements
$\vek{z}(\bar{\omega})$.

Bayesian update is based on the idea of Bayes' rule defined for
probabilities. Definition of Bayes' rule for continuous distribution
is, however, more problematic and hence \cite[Chapter
1.5]{Tarantola:2005} derived the posterior state of information
$\pi(\vek{\xi},\vek{z})$ as a conjunction of all information at hand
\begin{equation}
\pi(\vek{\xi},\vek{z}) = \kappa p_{\vek{\xi}}(\vek{\xi})
p_{\vek{z}}(\vek{z}) p(\vek{z}|\vek{\xi}) \, , \label{eq_posterior}
\end{equation}
where $\kappa$ is a normalization constant.

The posterior state of information defined in the space of model
parameters $\vek{\xi}$ is given by the marginal pdf
\begin{equation}
\pi_{\vek{\xi}}(\vek{\xi}) = \D{E}_{\mit{\bar{\Omega}}} \left[
\pi(\vek{\xi},\vek{z}) \right] =
%\int_{\mit{\bar{\Omega}}} \pi(\vek{q},\vek{z}) \, \di \bar{\omega} =
\kappa p_{\vek{\xi}}(\vek{\xi}) \int_{\mit{\bar{\Omega}}}
p(\vek{z}|\vek{\xi}) p_{\vek{z}}(\vek{z}) \, \D{P}(\di \bar{\omega})
= \kappa p_{\vek{\xi}}(\vek{\xi}) \, L(\vek{\xi}), \label{eq_margin-m}
\end{equation}
where $\mit{\bar{\Omega}}$ is a set of~random elementary events
$\bar{\omega}$ and measured data $\vek{z}$ enters through the {\it
  likelihood function} $L(\vek{\xi})$, which gives a measure of how good
a~numerical model is in explaining the data $\vek{z}$.

The most general way of extracting the information from the
posterior density $\pi_{\vek{\xi}}(\vek{\xi})$ is based on
sampling procedure governed by MCMC method. For more details about
this approach to Bayesian updating of uncertainty in description
of couple heat and moisture transport we refer to
\cite{Kucerova:2011:AMC}. In this paper, we focus on the
comparison of the posterior information obtained from the sampling
procedure using directly the computationally exhaustive numerical
model \eqref{eq:MS6} on one hand and using the PC approximation of
the model \eqref{eq_Tapprox} on the other hand.

\section{Numerical results for the Bayesian update}
\label{sec:bayes_res}

\begin{figure} [ht!]
\centering
\begin{tabular}{cc}
\includegraphics*[width=65mm,keepaspectratio]{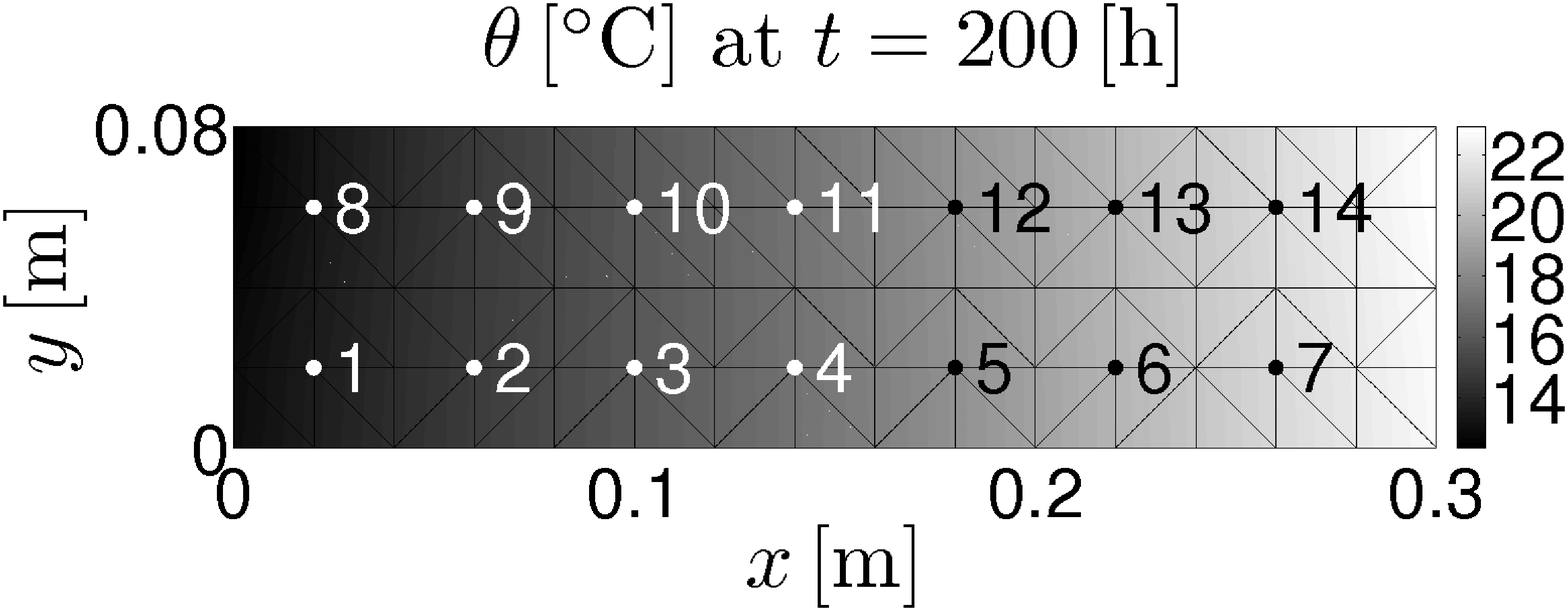}&
\includegraphics*[width=65mm,keepaspectratio]{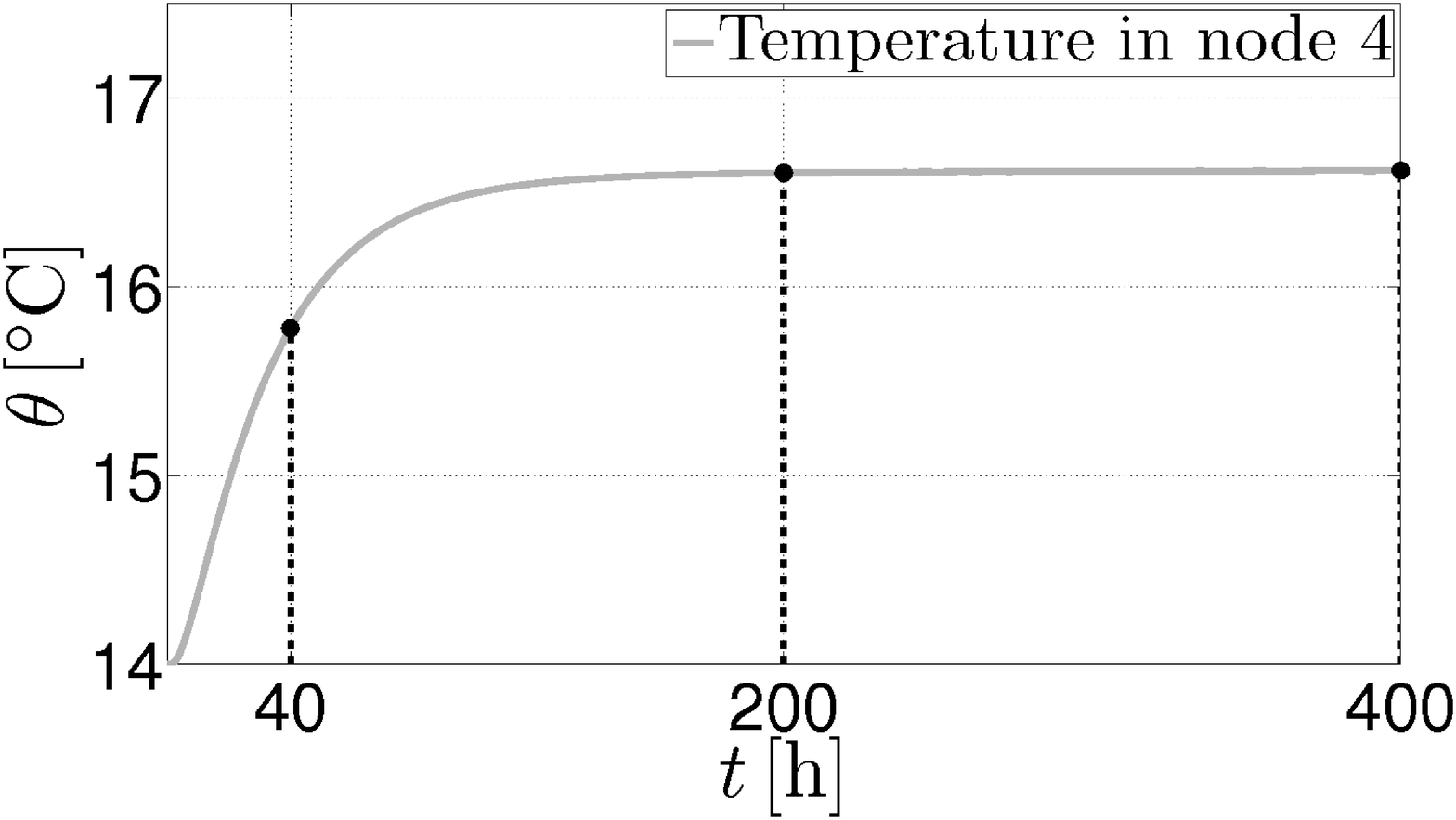}\\
(a)&(b)\\
\includegraphics*[width=65mm,keepaspectratio]{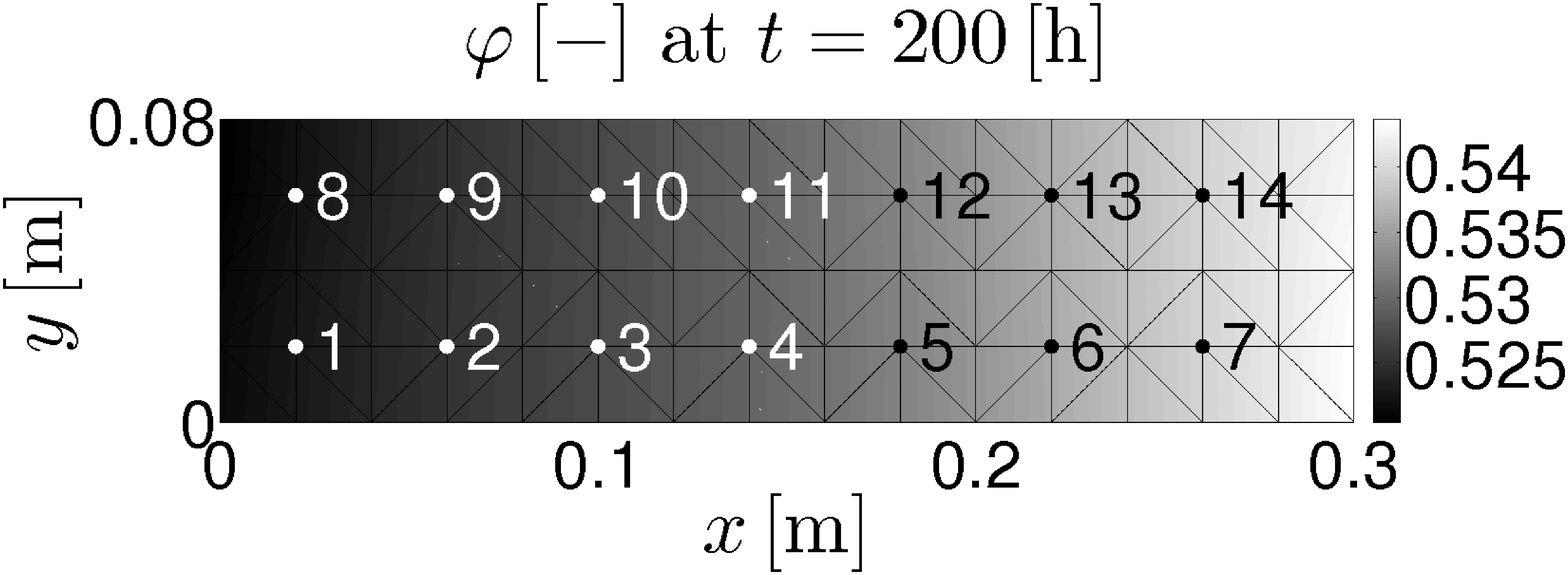}&
\includegraphics*[width=65mm,keepaspectratio]{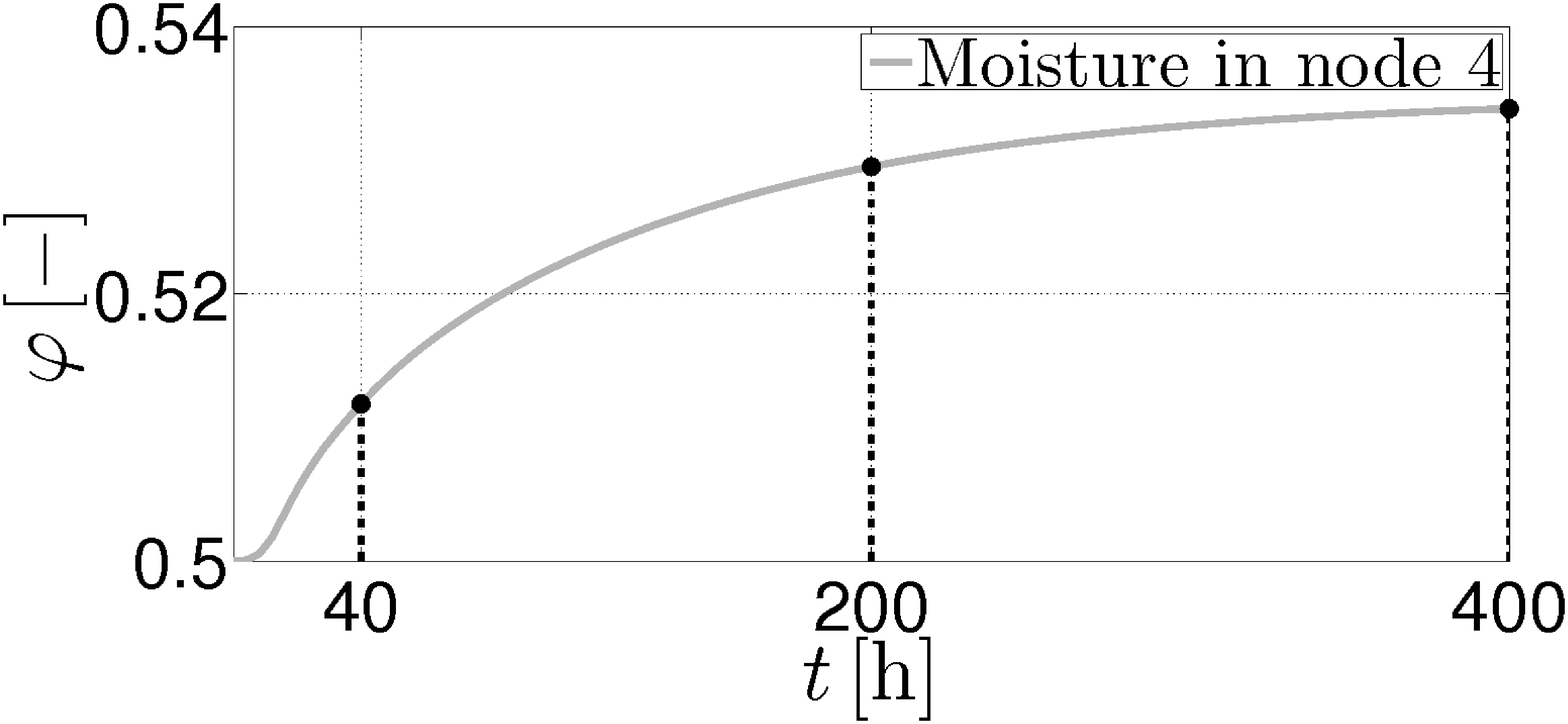}\\
(c)&(d)
\end{tabular}
\caption{Virtual observations: (a) and (c) spatial arrangement of
  probes; (b) and (d) temporal organization of measurements}
\label{fig_observations}
\end{figure}

Due to the lack of experimental data, we prepared a virtual
experiment using a FE simulation based on parameter fields
obtained by the KLE with $7$ eigenmodes so as to avoid the error
induced by KLE, which is mainly the subject of the work presented
in \cite{Kucerova:2011:AMC}. A related set of random variables
  $\vek{\xi}$ is drawn randomly from the prior distribution and stored
  for a purpose of latter comparison with the prior and the posterior
  state of knowledge. The resulting temperature and moisture fields
  considered as a so-called ``true state'' or simply the ``truth'' are
  shown in Fig.~\ref{fig_observations}. According to
\cite{Kucerova:2011:AMC} the values of temperature and moisture
are measured in 14 points (see Figs.~\ref{fig_observations} (a)
and (c)), and at three distinct times (see
Figs.~\ref{fig_observations} (b) and (d)). Hence, the observations
$\vek{z}$ consist of $84$ values.
To keep the presentation of the different numerical aspects of the
presented methods clear and transparent, we focus here on a quite
common and simple case, where modelling-uncertainties are
neglected and measurement errors are assumed to be Gaussian. Then
the likelihood function takes the form
\begin{equation}
  L({\vek{\xi}}) = \kappa\exp \left( - \frac{1}{2} \left(
      \vek{Y}({\vek{\xi}}) - \vek{z} \right)^{\mrm{T}} \mat{C}^{-1}_{\mrm{obs}}
    \left( \vek{Y}({\vek{\xi}}) - \vek{z} \right) \right) \, ,
\label{eq_likeli}
\end{equation}
where $\vek{Y}(\vek{\xi})$ is an observation operator mapping the
model response $\vek{u}$ given parameters $\vek{\xi}$ and loading
$\vek{f}$ to observed quantities $\vek{z}$. $\mat{C}_{\mrm{obs}}$
is a covariance matrix representing the uncertainty in
experimental error, which is obtained by perturbing the virtual
observations by Gaussian noise with standard deviation for
temperature $\sigma_{\theta} = 0.2$ $\mathrm{[^{\circ}C]}$ and for
moisture $\sigma_{\varphi} = 0.02$ $\mathrm{[-]}$ so as to get
$100$ virtual as an input for the covariance matrix evaluation. In
order to be able to compare the posterior state with the true
state also in terms of model parameters $\vek{\xi}$, we assume an
artificial situation where the observed quantities $\vek{z}$
correspond exactly to the true state of temperature and moisture.

\begin{figure} [ht!]
\centering
\begin{tabular}{cc}
\includegraphics*[width=65mm,keepaspectratio]{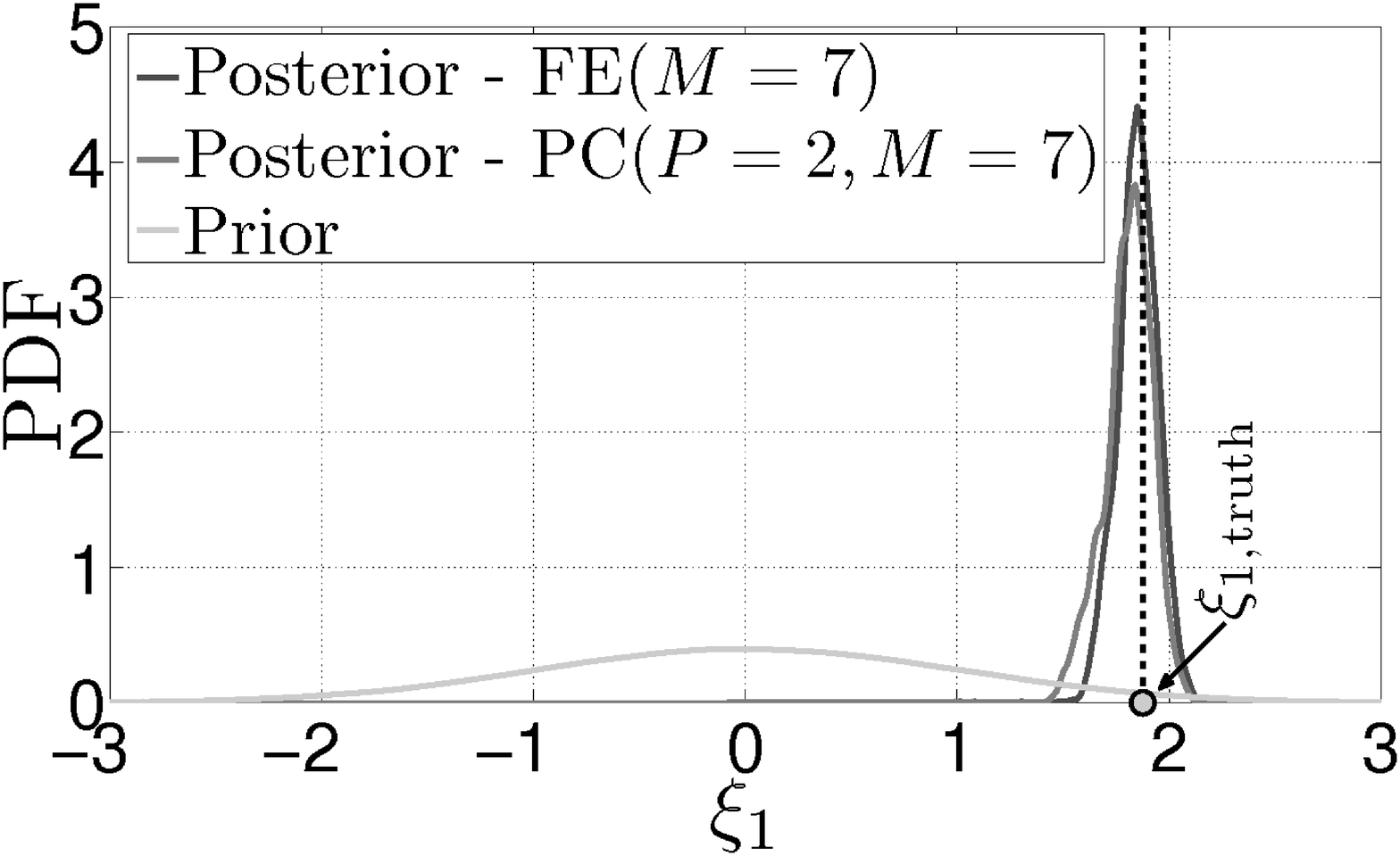}&
\includegraphics*[width=65mm,keepaspectratio]{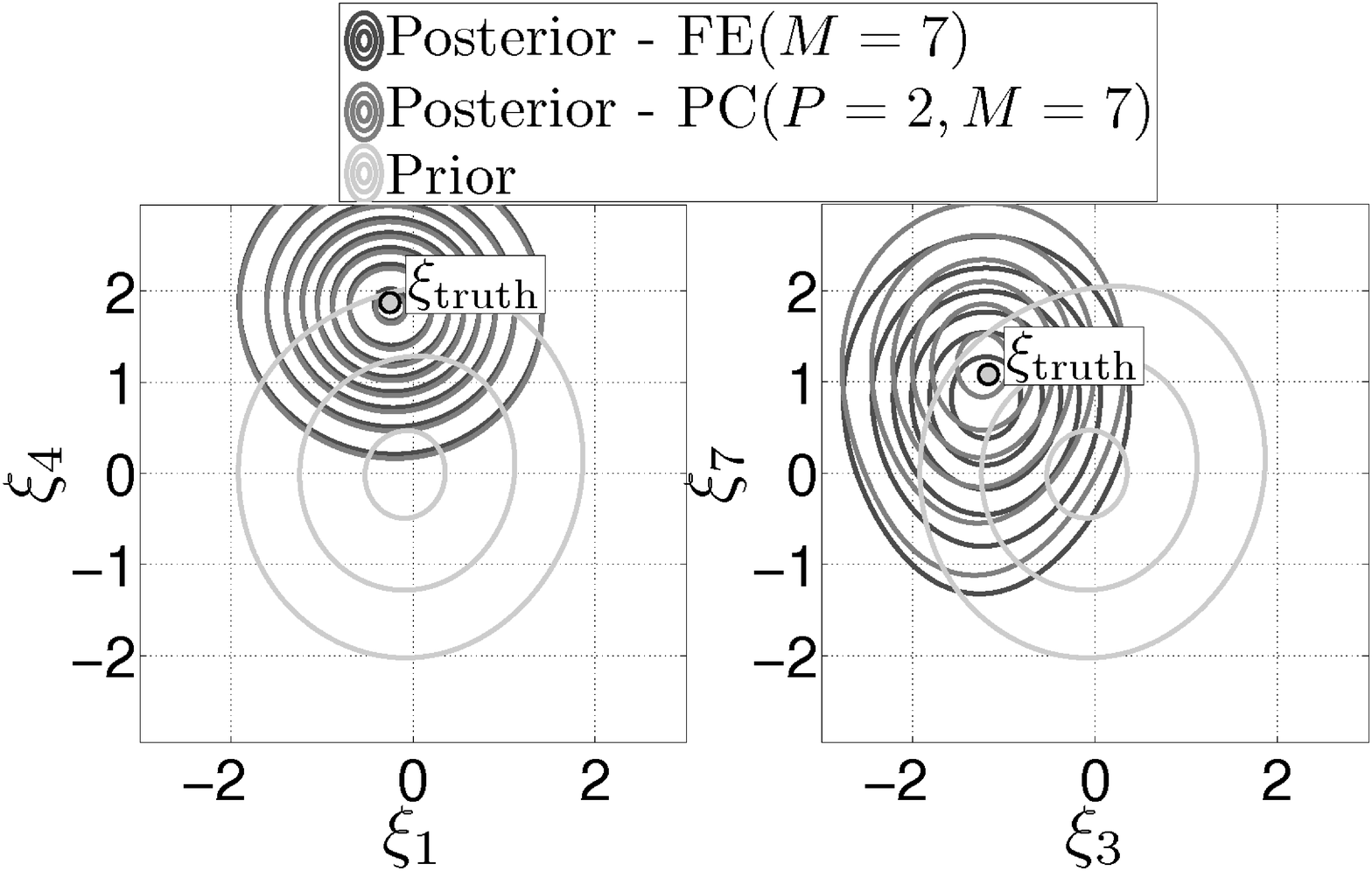}\\
(a)&(b)
\end{tabular}
\caption{Comparison of pdfs (a) for the separate variable $\xi_1$ and
  (b) for the pairs of variables.}
\label{fig_pdf_xi}
\end{figure}

The Bayesian update was performed using Metropolis-Hasting
algorithm and $100,000$ samples were generated in order to sample
the posterior density (\ref{eq_posterior}) over the variables
$\vek{\xi} = (\xi_1 \dots \xi_{M=7})$. The truth state, prior and
posterior pdfs obtained by the FE simulations and using the
PCE are plotted in Fig.~\ref{fig_pdf_xi}.
One can see that the error induced by PC surrogate of model
response are negligible in terms of the resulting posterior
densities. Figure~\ref{fig_pdf_xi} also demonstrates the fact that
the variables $\vek{\xi}$ being a priori standard Gaussian should
not be a posteriori Gaussian.

\begin{figure} [ht!]
\centering
\begin{tabular}{cc}
\includegraphics*[width=65mm,keepaspectratio]{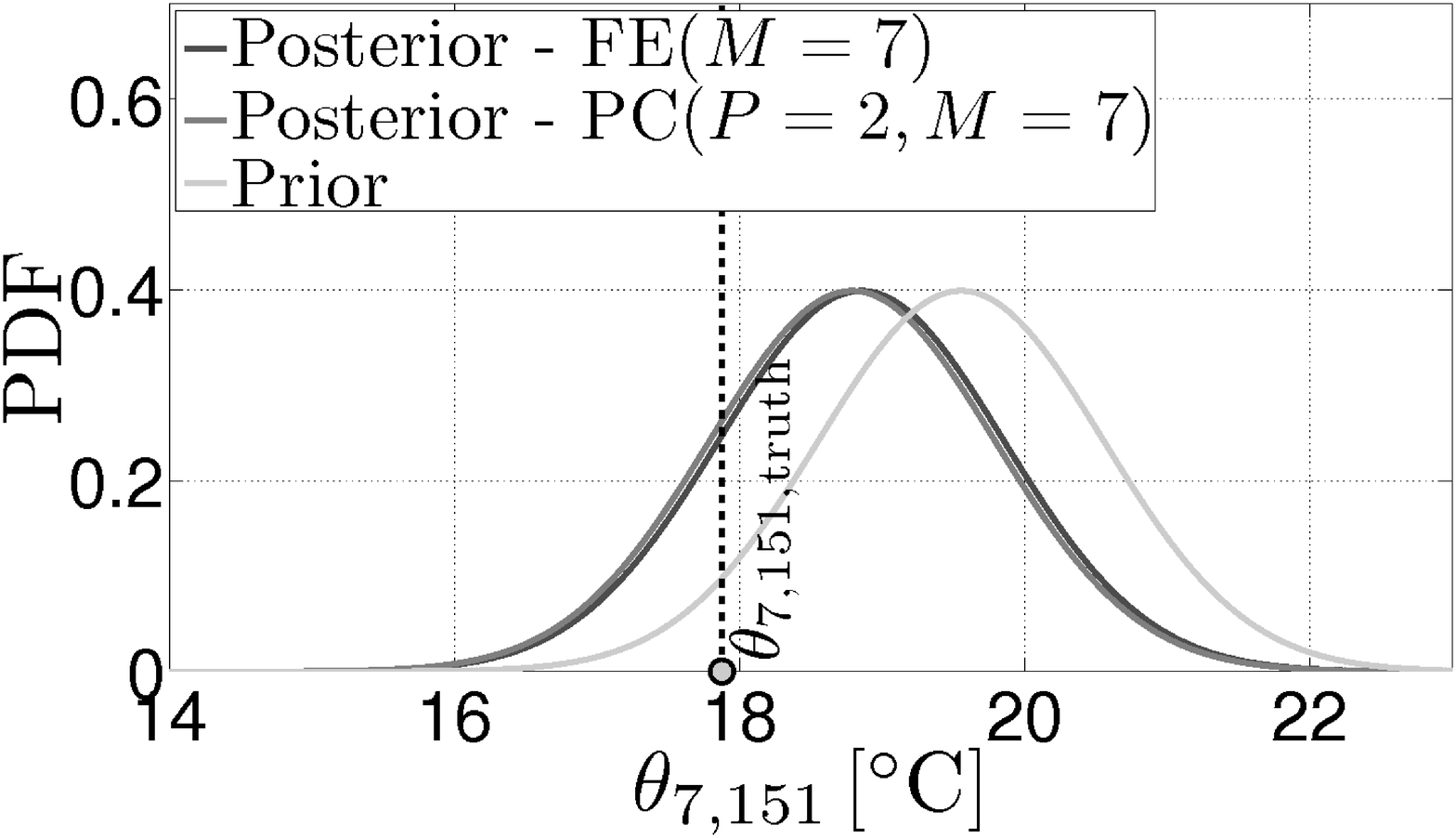}&
\includegraphics*[width=65mm,keepaspectratio]{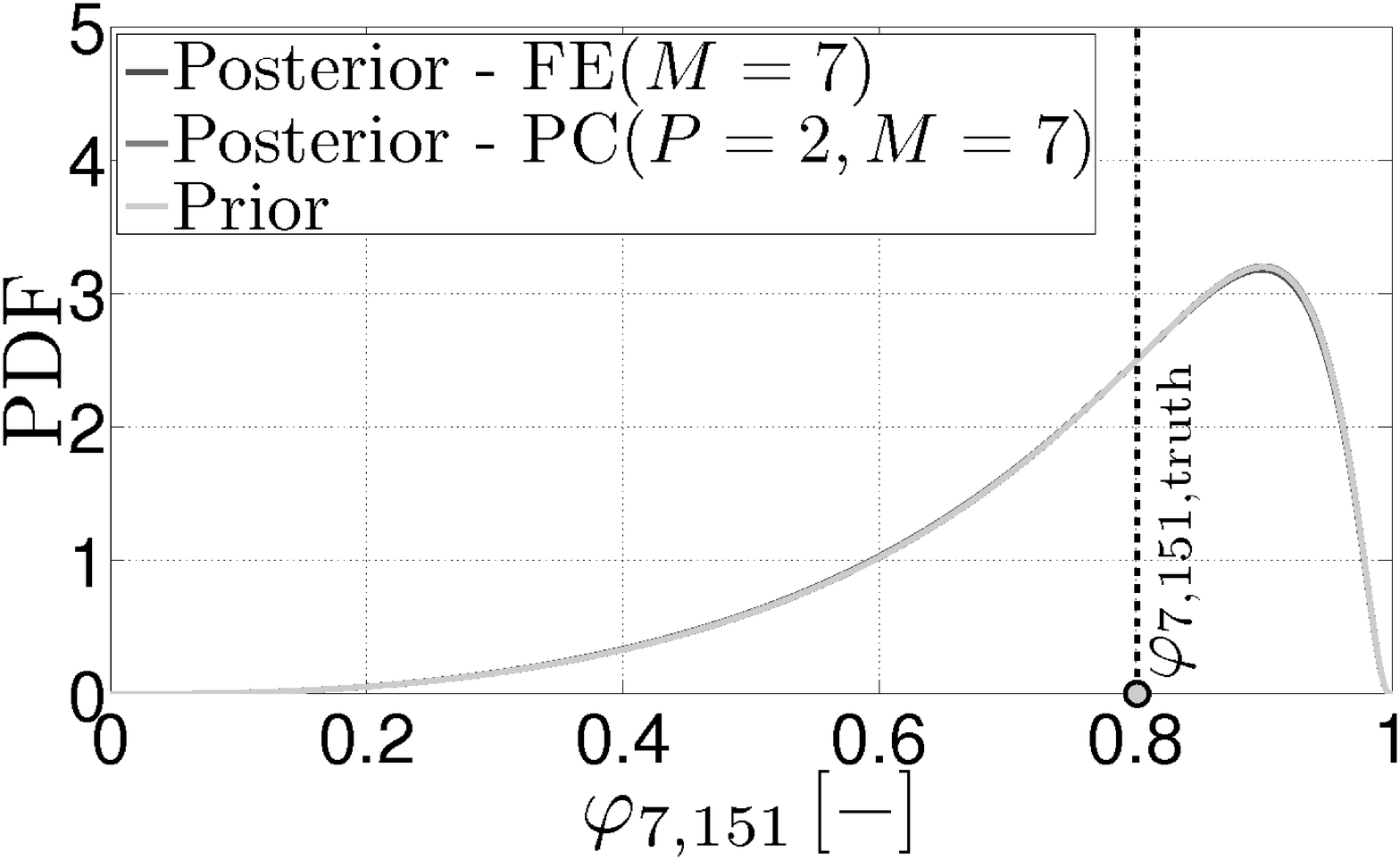}\\
(a)&(b)
\end{tabular}
\caption{comparison of pdfs (a) for the temperature and (b) for the
  moisture in FE node $7$ at $151$-th time step.}
\label{fig_pdf_resp}
\end{figure}

During the sampling procedure, we stored also the corresponding values
of parameter fields and response fields in order to obtain their
posterior state of information.
\begin{figure} [ht!]
\centering
\begin{tabular}{cc}
\includegraphics*[width=65mm,keepaspectratio]{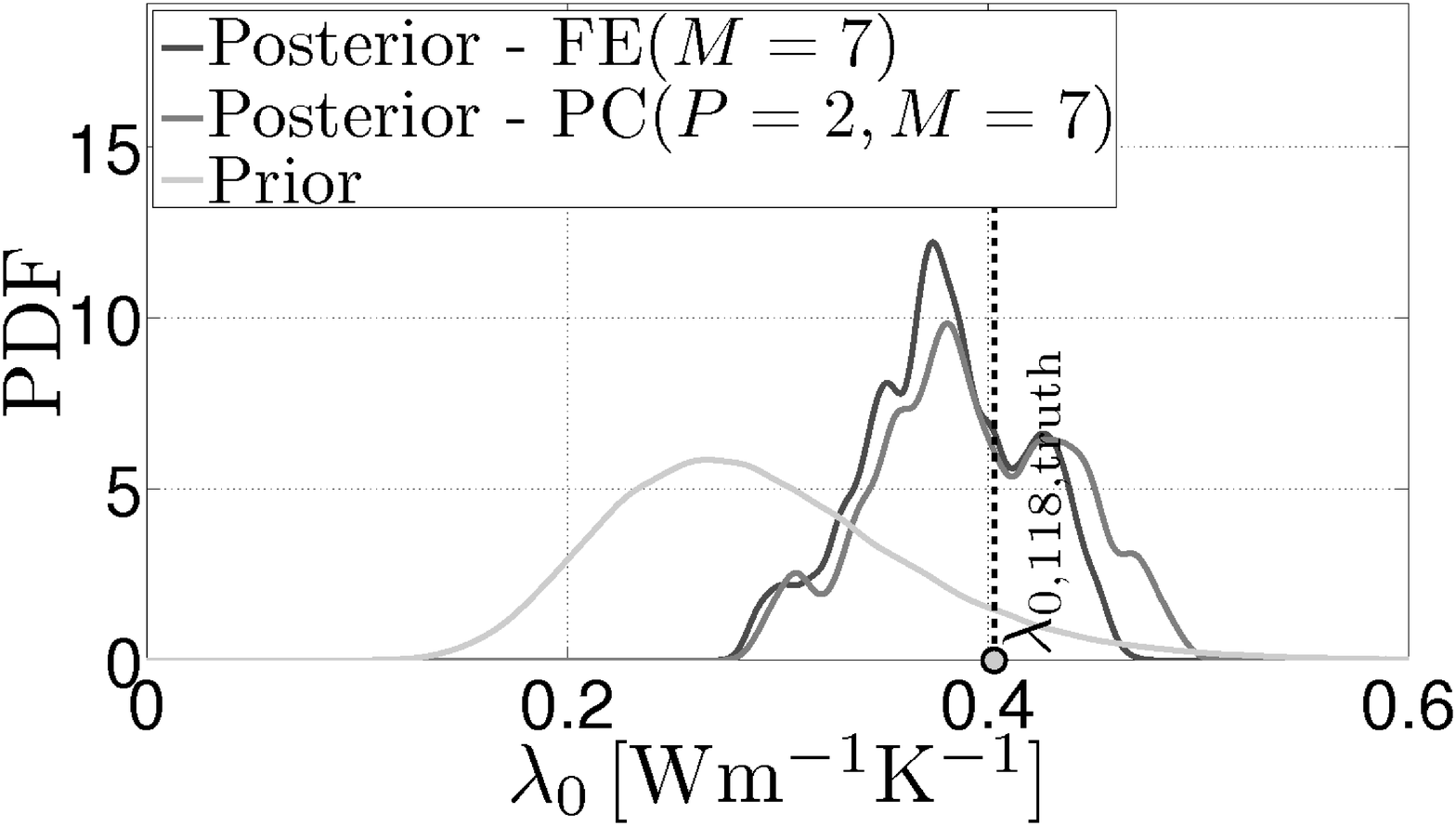}&
\includegraphics*[width=65mm,keepaspectratio]{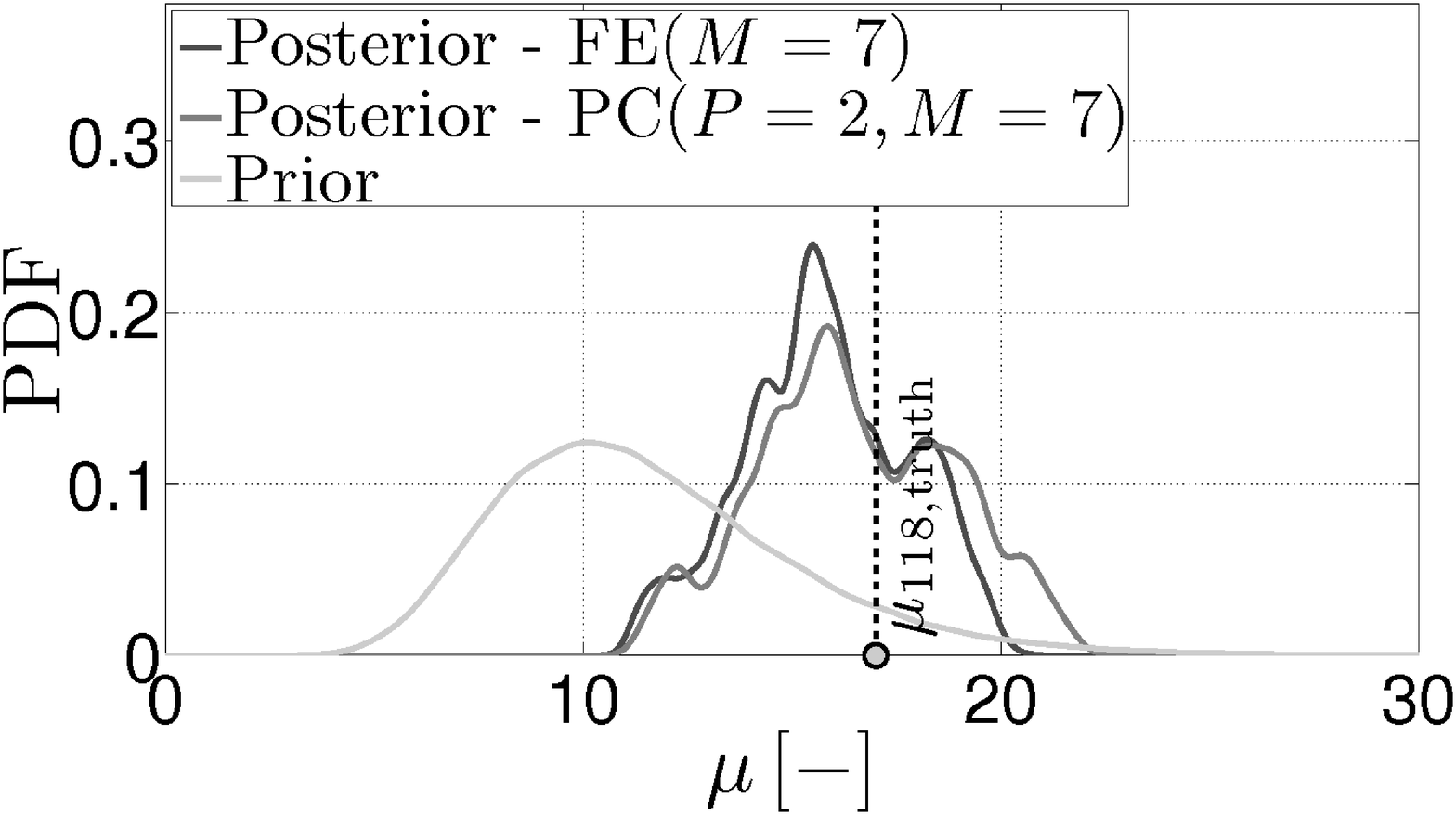}\\
(a)&(b)
\end{tabular}
\caption{Comparison of pdfs for material properties in FE node $7$:
  (a) the thermal conductivity of dry material $\lambda_0$ and (b)
  water vapour diffusion resistance factor $\mu$.}
\label{fig_pdf_matpar}
\end{figure}
As a result, Fig.~\ref{fig_pdf_matpar} shows the comparison of the
truth, and prior and posterior pdfs for two material parameters
$\lambda_0$ and $\mu$ in the top-right corner FE element, and
similarly, Fig.~\ref{fig_pdf_resp} presents pdfs for the
temperature and moisture in FE node $7$ at $400 [\mathrm{h}]$
(i.e. at the $151$-th time step).

We should note that the similarity of the prior and the posterior
pdfs for moisture in Fig.~\ref{fig_pdf_resp} is probably caused by
the very slight influence of the studied material parameters to
the moisture value or more precisely, the prior standard
deviations were very small.

\begin{figure} [ht!]
\centering
\includegraphics*[width=65mm,keepaspectratio]{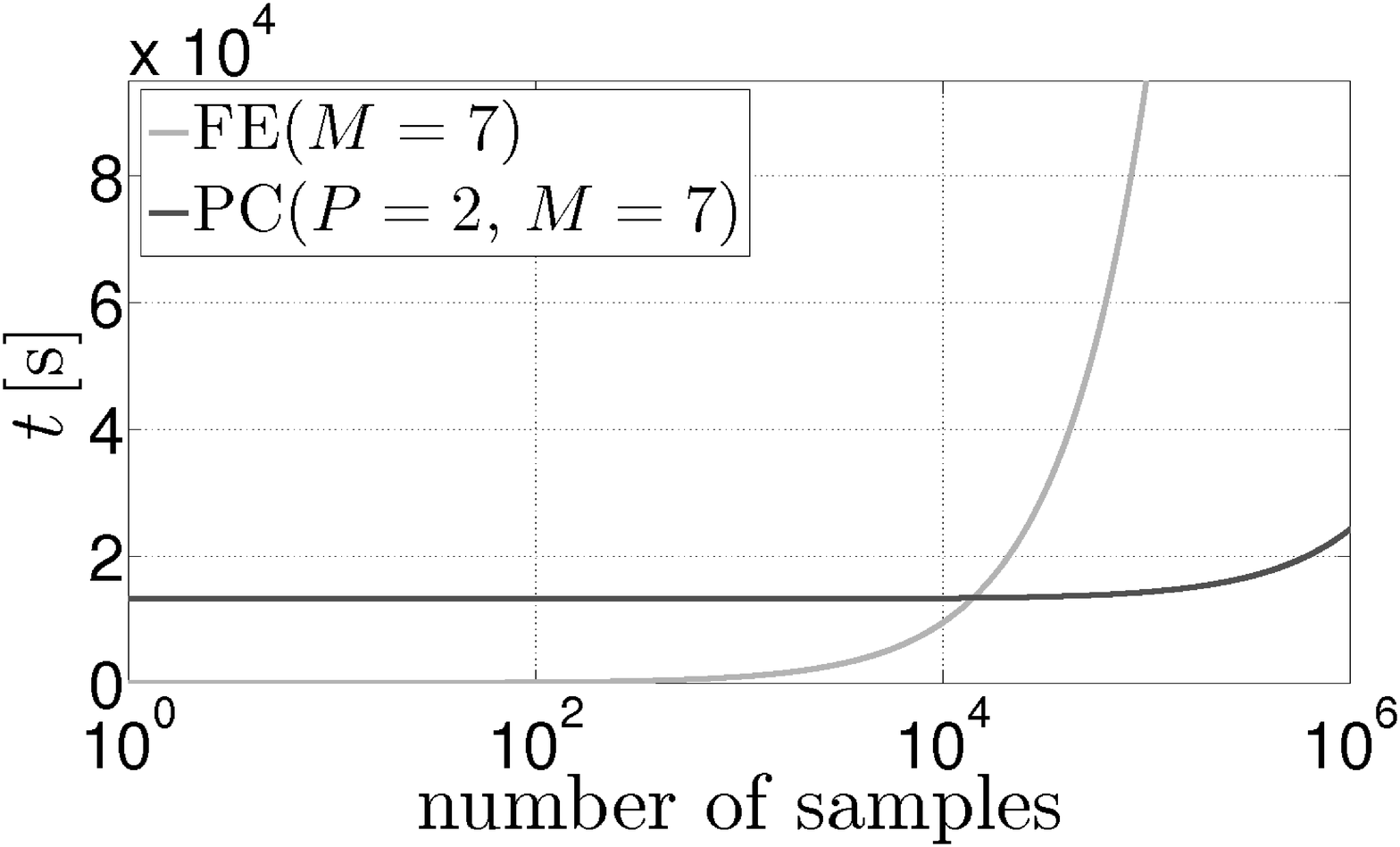}
\caption{Comparison of time necessary for evaluation of samples.}
\label{fig_timecomp}
\end{figure}

Beside the comparison of the PCE accuracy, we also compared the time
necessary to generate the samples. In case of PCE, the total time also
includes the time of PC coefficients computation. Particular
comparison of computational time needed by FE simulations and by PCE
evaluations is demonstrated in Fig.~\ref{fig_timecomp}.

\section{Conclusions}
\label{sec:concl}
The presented paper presents an efficient approach to propagation
and updating of uncertainties in description of coupled heat and
moisture transport in heterogeneous material. In particular, we
employed the K\"unzel's model, which is sufficiently robust to
describe real-world materials, but which is also highly nonlinear,
time-dependent and is defined by $8$ material parameters difficult
to be estimated from measurements. The updating procedure starts
with the prior information about the parameters' properties such
as positive-definitness and second order statistics. Heterogeneity
of the material under the study is taken into account by
describing the material properties by random fields, which are for
a simplicity considered as fully correlated. Then, the
corresponding correlation lengths are assume to be known as
another a priori information. In order to limit the number of
random variables necessary to describe the material, the random
fields are approximated by Karhunen-Lo\`eve expansion and hence,
all the remaining uncertainties are described by a set of standard
Gaussian variables whose number is given by the number of
eigenmodes involved in KLE.

These uncertainties are then propagated through the numerical
model so as to provide a probabilistic characterization of the
model response, here the moisture and temperature fields.
Simultaneously, the other information including uncertainties
coming from the experimental measurements is used to update the
prior uncertainties in the model parameters. In order to imitate
the experimental measurements, a virtual experiment is prepared
together with the relating uncertainties given by a covariance
matrix. The Markov Chain Monte Carlo method is then employed so as
to sample the posterior state of information.

The primary objective of the presented paper is to accelerate the
sampling procedure. To this goal, a polynomial chaos-based
approximation of the model response is constructed in order to
replace computationally expensive FE simulations by fast
evaluations of the PCE during the sampling. In particular, the PC
coefficients are obtained by an intrusive stochastic Galerkin
method. It is shown that the resulting approximations exhibit high
accuracy and the related posterior probability density functions
are sufficiently precise as well. Finally, the comparison of the
computational effort confirmed the large savings in case of PC
evaluations.

While the acceleration obtained by the presented procedure is
significant, it can be still unfeasible for very large problems.
Our future work will be focused on the elimination of the MCMC
sampling procedure itself by the update directly in terms of
parameters of probability density functions as proposed in
\cite{Rosic:2011:JCP}.

\section*{Acknowledgment}
This outcome has been achieved with the financial support of the
Czech Science Foundation, project No. 105/11/0411, the Czech
Ministry of Education, Youth and Sports, projects No.
MSM6840770003 and No. MEB101105 and the German Research Foundation
(DFG) project No. MA 2236/14-1.

\bibliographystyle{elsarticle-num}
\bibliography{liter_arxiv}

\end{document}